\documentclass[12pt]{article}
\usepackage{mathtools}
\usepackage{amsmath}
\usepackage{amsfonts}
\usepackage{dsfont}
\usepackage{amssymb}
\usepackage{graphicx}
\graphicspath{ {./images/} }
\usepackage{natbib}
\setcitestyle{authoryear,open={(},close={)},maxcitenames=2}
\usepackage[all]{xy}
\usepackage{xcolor}
\usepackage{multirow}
\usepackage{hyperref}
\usepackage[]{algorithm2e}
\usepackage{cleveref}
\usepackage{tikz}

\DeclareMathAlphabet{\mathcalligra}{T1}{calligra}{m}{k}
\DeclareMathAlphabet{\mathpzc}{OT1}{pzc}{m}{it}

\date{}

\begin{document}
\begin{titlepage}

\begin{center}
\bf\LARGE 

Reinforcement Learning\\in Economics and Finance
\par\end{center}

\bigskip

\begin{center}
\Large by 
\par\end{center}

\renewcommand*{\thefootnote}{\fnsymbol{footnote}}
\begin{center}
\Large
\bigskip
\textbf{Arthur Charpentier}\\[1ex]
\large
Universit\'e du Qu\'ebec \`a Montr\'eal (UQAM)\\
201, avenue du Pr\'esident-Kennedy, \\
Montr\'eal (Qu\'ebec), Canada H2X 3Y7
\\ arthur.charpentier{@}uqam.ca

\Large
\bigskip
\textbf{Romuald \'Elie}\\[1ex]
\large
LAMA, Universit\'e Gustave Eiffel, CNRS \\
5, boulevard Descartes\\
Cité Descartes - Champs-sur-Marne\\
77454 Marne-la-Vallée cedex 2, France
\\ romuald.elie@u-pem.fr

\Large
\bigskip
\textbf{Carl Remlinger}\\[1ex]
\large
LAMA, Universit\'e Gustave Eiffel, CNRS \\
5, boulevard Descartes\\
Cité Descartes - Champs-sur-Marne\\
77454 Marne-la-Vallée cedex 2, France

\par\end{center}

\setcounter{footnote}{0}

\vfill
\begin{center}
\large 
March 2020
\par\end{center}

\vfill
%\indent Flachaire acknowledges the support of the Institut Universitaire de France and the A*MIDEX project (ANR-11-IDEX-0001-02) funded by the "Investissements d'Avenir" French Government program, managed by the French National Research Agency (ANR) @@

\end{titlepage}

\begin{abstract}
Reinforcement learning algorithms describe how an agent can learn an optimal action policy in a sequential decision process, through repeated experience. In a given environment, the agent policy provides him some running and terminal rewards. As in online learning, the agent learns  sequentially. As in multi-armed bandit problems, when an agent picks an action, he can not infer ex-post the rewards induced by other action choices. In reinforcement learning, his actions have consequences: they influence  
%whose machine will have to take actions. Those actions will have consequences, and will influence 
not only rewards, but also future states of the world. The goal of reinforcement learning is to find an optimal policy -- a mapping from the states of the world to the set of actions, in order to maximize cumulative reward, which is a long term strategy. Exploring might be sub-optimal on a short-term horizon but could lead to optimal long-term ones. Many problems of optimal control, popular in economics for more than forty years, can be expressed in the reinforcement learning framework, and recent advances in computational science, provided in particular by deep learning algorithms, can be used by economists in order to solve complex behavioral problems. In this article, we propose a state-of-the-art of reinforcement learning techniques, and present applications in economics, game theory, operation research and finance.
\end{abstract}



\RestyleAlgo{boxruled}

%%%%%%%%%%%%%%%%%%%%%%%%%%%%%%%%%%%%%%%%%%%%%%%%%%%%
%\maketitle
%\input{abstract.tex}

%%%%%%%%%%%%%%%%%%%%%%%%%%%%%%%%%%%%%%%%%%%%%%%%%%%%
%\newpage
%\textcolor{red}{Carl: l'ancien main.tex est dans mains/original_main.tex}\\
%\textcolor{red}{Carl: the table of content will not be kept}
%\tableofcontents

%%%%%%%%%%%%%%%%%%%%%%%%%%%%%%%%%%%%%%%%%%%%%%%%%%%%
\newpage
\section{Introduction}
% \carl{
% 1. data: toute les données (supervisé) -> données sequentielles (online) -> crée les données (bandits,rl)\\ 
% 2. info disponibles, rejouer le match ou pas\\
% 3. NewEstimate = OldEstimate + StepSize(Target - OldEstimate)\\
% }

\subsection{An Historical Overview}

Reinforcement learning is related to the study of how agents, animals, autonomous robots use experience to adapt their behavior in order to maximize some rewards. It differs from other types of learning (such as unsupervized or supervised) since the learning follows from feedback and experience (and not from some fixed training sample of data). \citet{thorndike} or \citet{skinner1938behavior} used reinforcement learning in the context of behavioral psychology, ethology and biology. For instance, \citet{thorndike} studied learning behavior in cats, with some popular experiences, using some `puzzle box' that can be opened (from the inside) via various mechanisms (with latches and strings) to obtain some food that was outside the box. Edward Thorndike observed that cats usually began experimenting -- by pressing levers, pulling cords, pawing, etc. -- to escape, and over time, cats will learn how particular actions, repeated in a given order, could lead to the outcome (here some food). To be more specific, it was necessary for cats to explore alternative actions in order to escape the puzzle box. Over time, cats did explore less, and start to exploit experience, and repeat successful actions to escape faster. And the cat needed enough time to explore all techniques, since some could possibly lead more quickly -- or with less effort -- to the escape. \citet{thorndike} proved that there was a balance between exploration and exploitation.  This issue could remind us of the {\em simulated annealing} in optimization, where a classical optimization routine is pursued, and we allow to move randomly to another point (which would be the exploration part) and start over (the exploitation part). Such a procedure reinforces the chances of converging towards a global optimum, instead of converging to a more local one.

Another issue was that a multi-action sequence was necessary to escape, and therefore, when the cat was able to escape at the first time it was difficult to assign which action actually caused the escape. An action taken at the beginning (such as pulling a string) might have an impact some time later, after other actions are performed. This is usually called a credit assignment problem, as in \citet{Minsky1961}. \citet{skinner1938behavior} refined the puzzle box experiment, and introduced the concept of {\em operant conditioning} (see \citet{jenkins} or \citet{garcia_1981} for an overview). The idea was to modify a part, such as a lever, such that at some points in time pressing the lever will provide a positive reward (such as food) or a negative one (i.e. a punishment, such as electric shocks). The goal of those experiments was to understand how past voluntary actions modify future ones. Those experiments were performed on rats, and no longer cats. \citet{tolman1948cognitive} used similar experiments (including also mazes) to prove that the classical approach, based on chaining of stimulus-responses, was maybe not the good one to model animal (and men) behaviors. A pure stimulus-responses learning could not be used by rats to escape a maze, when experimenters start to block roads with obstacles. He introduced the idea of {\em cognitive maps} of the maze that allow for more flexibility. All those techniques could be related to the ones used in reinforcement learning.

Reinforcement learning is about understanding how agents might learn to make optimal decisions through repeated experience, as discussed in \citet{Sutton81}. More formally, agents (animals, humans or machines) strive to maximize some long-term reward, that is the cumulated discounted sum of future rewards, as in classical economic models. Even if animals can be seen as have a short-term horizon, they do understand that a punishment followed by a large reward can be better than two small rewards, as explained in \citet{Rescorla79}, that introduced the concept of second-order conditioning. A technical assumption, that could be seen as relevant in many human and animal behaviors, is that the dynamics satisfies some Markov property, and in this article we will focus only on Markov decision processes. Reinforcement learning is about solving the credit assignment problem by matching actions, states of the world and rewards.

\begin{figure}[!ht]
    \centering
    \includegraphics[width=.9\textwidth]{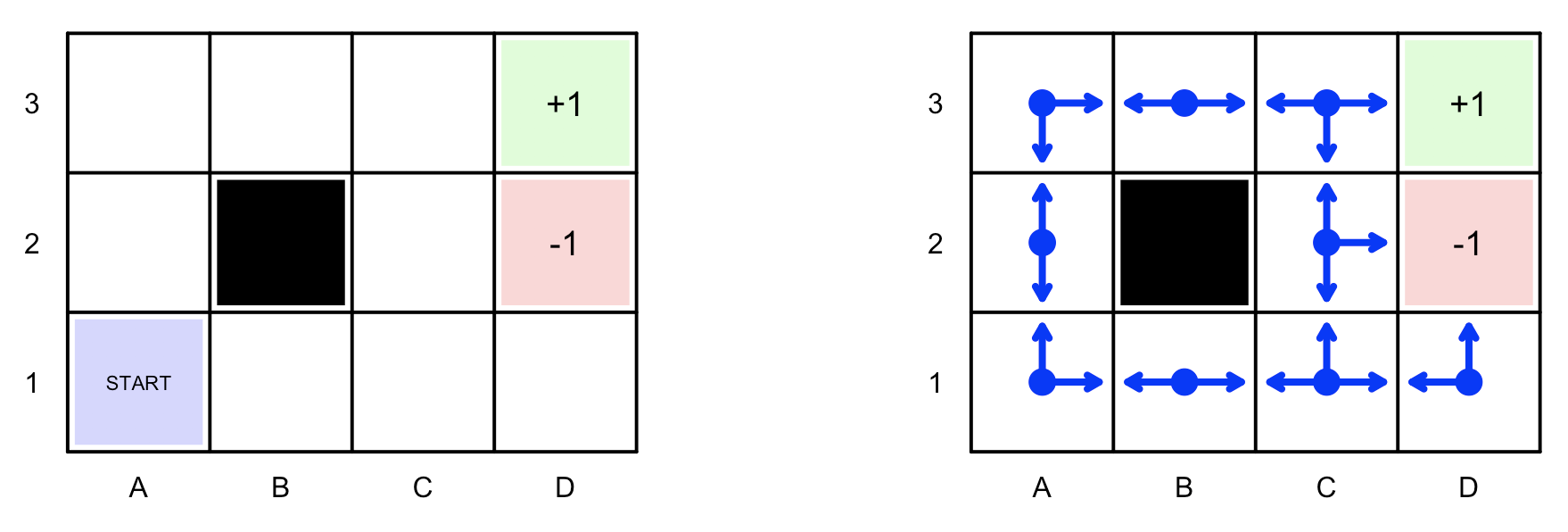}\\
    \includegraphics[width=.9\textwidth]{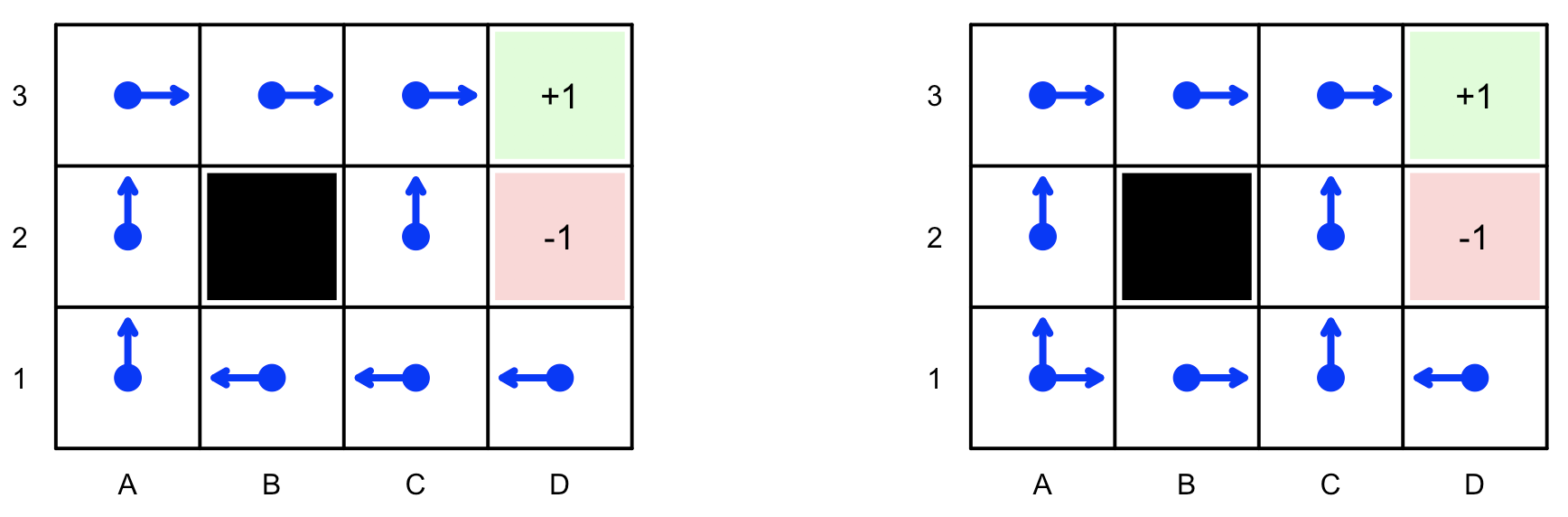}
    \caption{Sequential decision making problem on a $4\times3$ grid ($\mathcal{S}$ states), from \citet{Russell:2009}. The agent starts at the state ({\sffamily A,1}), and moves around the environment, trying to reach terminal state ({\sffamily D,3}) to get a +1 reward - and to avoid terminal state ({\sffamily D,2}) where a -1 reward (punishment) is given. Possible actions ($\mathcal{A}$) are given on the top-right figure. On the bottom, two policies are given with $\pi:\mathcal{S}\rightarrow\mathcal{A}$ on the left, and $\pi:\mathcal{S}\rightarrow A\subset\mathcal{A}$ on the right. In the later case, there can be random selection of actions in some states, for instance $\pi(\text{({\sffamily A,1})})\in\lbrace\text{up},\text{right}\rbrace$.}
    \label{fig:Russell:2}
\end{figure}

As we will see in the next section, formally, at time $t$, the agent at state of the world $s_{t}\in\mathcal{S}$ makes an action $a_t\in\mathcal{A}$, obtains a reward $r_t\in\mathcal{R}$ and the state of the world becomes $s_{t+1}\in\mathcal{S}$. A policy is a mapping from $\mathcal{S}$ to $\mathcal{A}$, and the goal is to learn from past data (past actions, past rewards) how to find an optimal policy. A popular application of reinforcement learning algorithms is in games, such as playing chess or Go, as discussed in \citet{Silver1140}, or \citet{igami2017artificial} which provides economic interpretation of several algorithms used on games (Deep Blue for chess or AlphaGo for Go) based on structural estimation and machine (reinforcement) learning. More simply, \citet{Russell:2009} introduced a grid world to explain heuristics about reinforcement learning, see Figure \ref{fig:Russell:2}. Positions on the $4\times3$ grid are the states $\mathcal{S}$, and actions $\mathcal{A}$ are movements allowed.
The optimal policy $\pi:\mathcal{S}\rightarrow\mathcal{A}$ is here computed using sequential machine learning techniques that we will describe in this article.

\subsection{From Machine to Reinforcement Learning}

Supervised Machine Learning techniques is a static problem: given a dataset $\mathcal{D}_n=\{(y_i,{x}_i)\}$, the goal is to learn a mapping $\widehat{m}_n$ between ${x}$ and $y$. In decision theory $\widehat{m}_n$ typically takes values in a binary space, which could be to accept or reject a mortgage in credit risk models, or to invest or not in some specific asset. $\widehat{m}_n$ can also take values in the real line, and denote an amount of money to save, a quantity to purchase or a price to ask. Online learning is based on the assumption that $(y_i,{x}_i)$ arrive in a sequential order, and the focus is on the evolution of $\widehat{m}_n$ as $n$ growth, updating the training dataset from $\mathcal{D}_{n-1}$ to $\mathcal{D}_n$. Reinforcement learning incorporates the idea that at time $n-1$, a choice was made, that will influence $(y_n,{x}_n)$, and the standard i.i.d. assumption of the dataset is no longer valid. Reinforcement learning is related to sequential decision making and control. 

Consider an online shop, where the retailer tries to maximize profit by sequentially suggesting products to consumers. Consumers are characterized by some features, such as their age, or their gender, as well as information about what's in their shopping cart. The consumer and the shop will have  sequential interactions. Each round, the consumer can either add a product to the shopping cart, or not buy a product and continue shopping, or finally stop shopping and check out. Those transitions are characterized by transition probabilities, function of past states and actions. Such transition probability function is unknown and must be learned by the shop. Should the retailer display the most profitable products, exploiting information he obtained previously, or explore actions, that could be less profitable, but might provide relevant information ?

The induced problems are related to the fact that acting has consequences, possibly delayed. It is about learning to sacrifice small immediate rewards in order to gain larger long-term ones. If standard Machine Learning is about learning from given data, reinforcement learning is about active experimentation. Actions can be seen as an intervention, so there are strong connections between reinforcement learning and causality modeling. Reinforcement learning allows us to infer consequences of interventions (or actions) used in the past. \cite{Pearl2019} asked the simple economic question `{\em what will happen if we double the price}' (of an item we try to sell)? `{\em Such questions cannot be answered from sales data alone, because they involve a change in customers behaviour, in reaction to the new pricing}'. Reinforcement learning is related to such problem: inferring the impact of interventions. And the fact that intervention will impact the environment, mentioned by \cite{Pearl2019}, is precisely what reinforcement learning is about. So this theory, central in decision science will appear naturally in sequential experimentation, optimization, decision theory, game theory, auction design, etc. As we will see in the article (and as already mentioned in the previous section), models in sequential decision making as long history in economics, even if rarely mentioned in the computational science literature. Most of the articles published in economic journal mentioned that such problems were computationally difficult to solve. Nevertheless, we will try to show that recent advances are extremely promising, and it is now to possible to model more and more complex economic problems.

\subsection{Agenda}

In section \ref{sec:ML2RL}, we will explain connections between {\em reinforcement learning} and various related topics. We will start with {\em machine learning} principles, defining standard tools that will be extended later one (with the loss function, the risk of an estimator and regret minimization), in section \ref{sec:machinelearning}. In section \ref{sec:online}, we introduce dynamical problems with {\em online learning}, where we exploit past information sequentially. In section \ref{sec:bandits}, we present briefly the {\em multi-armed bandit} problem, where choices are made, at each period of time, and those have consequences on the information we obtain. And finally, in section \ref{sec:reinforcement} we start formalizing {\em reinforcement learning} models, and give a general framework. In those sections, we mainly explain the connections between various learning terms used in the literature.

Then, we present various problems tackled in the literature, in section \ref{sec:desc}. We will start with some general mathematical properties, giving various interpretations of the optimization problem, in section \ref{sec:maths}. %In section \ref{sec:model-based}, we will start with the model-based approach, while the model-free approaches will be discussed afterwards, in section \ref{sec:model-free-TD} with the temporal-difference approach, and with approximate solution methods in section \ref{sec:approx}. 
Finally, we will conclude, in section \ref{sec:inverse}, with a presentation of a classical related problem, called inverse reinforcement learning, where we try to use observed decisions in order to infer various quantities, such as the reward or the policy function.

Finally,  three sections are presenting applications of reinforcement learning. In section \ref{sec:econ}, we discuss applications in economic modeling, starting with the classical consumption and income dynamics, which is a classical optimal control problem in economics. We then discuss bounded rationality and strong connections with reinforcement learning. Then we will see, starting from \citet{Jovanovic82}, that reinforcement learning can be used to model single firm dynamics. And finally, we present connections with adaptative design for experiments, inspired by \citet{weber1992} (and multi-armed bandits).

In section \ref{sec:games}, we discuss applications of reinforcement learning in operation research, such as the traveling salesman, where the standard dilemma exploration/exploitation can be used to converge faster to (near) optimal solutions. Then we discuss stochastic games and equilibrium, as well as mean-field games, and auctions and real-time bidding. Finally, we will extend the single firm approach of the previous section to the case of oligopoly and dynamic games.

Finally, in section \ref{sec:finance}, we detail applications in finance. We start with risk management, valuation and hedging of financial derivatives problems on then focus on portfolio allocation issues. At last, we present a very natural framework for such algorithms: market impact and market making.

%%%%%%%%%%%%%%%%%%%%%%%%%%%%%%%%%%%%%%%%%%%%%%%%%%%%
%\newpage
\section{From Machine to Reinforcement Learning}\label{sec:ML2RL}
%\textcolor{red}{Carl: maybe include a "How to Explain the Prediction of a Machine Learning Model?" part}\\

Machine learning methods generally make decision based on known properties learned from the training data, using many principles and tools from statistics. However machine learning models aspire to find generalized predictive pattern. Most learning problems could be seen as an optimization of a cost: minimizing a loss or maximizing a reward. But learning algorithms seek to optimize a criterion (loss, reward, regret) on training {and} unseen samples. 

\subsection{Machine Learning principles}\label{sec:machinelearning}
Machine learning has so many branches (supervised vs unsupervised learning, online or not,...) that it is not always easy to identify the label associated to a given real world problem. Therefore, seeing machine learning as a set of data and an optimization criterion is often helpful.
To introduce Reinforcement Learning (RL), we propose here a regret approach, which ties machine learning, online aggregation, bandits and, more generally, reinforcement learning.

%\subsubsection{Supervised Learning}
In order to introduce most of machine learning terminology and schemes, we detail a class of models: supervised learning. In this class of models, one variable is the variable of interest, denoted $y$ and usually called the endogeneous variable in econometrics.
To do so, consider some learning sample $\mathcal{D}_n=\{(y_1,x_1), ..., (y_n,x_n)\}$ seen as realization of $n$ i.i.d. random variables $(Y,X)$. 
We wish to map the dataset $\mathcal{D}_n$ into a model from the (supposed) statistical relations between $x_i$ and $y_i$ that are relevant to a task. Note that in the context of sequential data we will prefer the generic notation $(y_t,x_t)$.

The goal, when learning, is to find a function $f\in\mathcal{F}$ from the input space $\mathcal{X}$ into the action space $\mathcal{A}$:
$f: \mathcal{X} \mapsto \mathcal{A}$.
Thus, $f(x|\mathcal{D}_n)$ is the action at some point $x$. % and $f(x|D_n)$ is the action seen as a random variable. Indeed, $\mathcal{D}_n$ is a realization of $D_n=\{(Y_i,X_i)\}$.
An action could be a prediction (for example what temperature will it be tomorrow? Is there a cat on this image?) or a decision (a chess move, go move...).
Note that in a standard regression problem $\mathcal{A}$ is the same as $\mathcal{Y}$, but not necessary in a classification problem: in a logistic regression, $\mathcal{Y}=\{0,1\}$ but actions can be probabilities $\mathcal{A}\in[0,1]$.

The decision function $f$ is all the better as its actions $f(x)$ are good when confronted to the unseen corresponding output $y$ from $\mathcal{Y}$. The loss function (or cost) measures the relevance of these actions when $f(x)$ is taken and $y$ has occurred:
$\ell: \mathcal{A} \times \mathcal{Y} \mapsto \mathbb{R}_+$.

The risk is the expectation of the loss: 
$$
\mathcal{R}(f) = \mathbb{E}\big[\ell(f(X), Y)\big]
$$
Thus formalized, the learning could be seen as an optimization problem. We wish to find a function $f^*\in\mathcal{F}$ which minimizes the cost: $$\mathcal{R}(f^*) = \inf_{f\in\mathcal{F}}\left\lbrace\mathcal{R}(f)\right\rbrace$$
If such a function $f^*$ exists and is unique it is called oracle or target.

In most applications we do not know the distribution of the data. However, given a training set $\mathcal{D}_n=\{(x_1,y_1),\dots,(x_n,y_n)\}$, we use the empirical distribution of the training data and define 
$$\widehat{\mathcal{R}}_n(f)=\frac{1}{n}\sum_{i=1}^n \ell(f(x_i),y_i).$$
Thus, we minimize this empirical risk while trying to avoid over-fitting and keeping in mind that the real objective is to minimize $\mathcal{R}(f)$, i.e. the average loss computed on any new observation. The main difficulty is that the target function is only defined at the training points.

Furthermore, we need to restrain the class of target functions or loss function class. Indeed, It would be impossible to reach sub-linear regret: if the loss is bounded $0\leq \ell\leq K$ then $\mathcal{R}_n\leq Kn$, hopefully $\mathcal{R}_n \ll n$

% \carl{point sur biais-variance?}\\
% \textbf{Risk decomposition}\\
% Called approximation-estimation trade-off or bias-variance trade-off
% \begin{eqnarray}
% \big(\mathcal{R}(\widehat{f_S}) - \mathcal{R}(f^*) \big) = \big(\mathcal{R}(\widehat{f}_S) - \mathcal{R}(f_S^*) \big) + \big( \mathcal{R}(f_S^*) - \mathcal{R}(f^*) \big)
% \end{eqnarray}
% The first left term is \textbf{excess risk}, the first right term is\textbf{ estimation error} and the second right term is the \textbf{approximation error}. blabla\\

% \begin{figure}[h!]
%   \begin{center}
%   \includegraphics[width=0.8\textwidth]{biasvariancetradeoff.png}
%   \end{center}
%   \label{fig:biasvariance}
%   \caption{Bias-variance trade-off}
% \flushleft{\textit{blabla. Note: a recent research have shown that generalization error can decrease if multiply parameters (increase capacity) and explore to the right the over-fitting zone. It would be an explanation why neural networks give such impressive results.}}
% \end{figure}

%\carl{point sur la regularization?}\\

One way to evaluate the learning performance is to compute regret. Regret is defined as the difference between the actual risk, and the optimal oracle risk,
\begin{eqnarray}
R &=& \mathcal{R}(f) - \mathcal{R}(f^*) \nonumber\\
               &=& \mathcal{R}(f) - \inf_{f\in\mathcal{F}}\left\lbrace\mathcal{R}(f)\right\rbrace \nonumber\\
               &=& \mathbb{E}\big[\ell(f(X),Y)\big] - \mathbb{E}\big[\ell(f^*(X),Y)\big]. \nonumber
\end{eqnarray}
In supervised learning, we prefer the name of excess risk, or excess loss. This notion of regret is particularly relevant in sequential learning, where your action at $t$ depends on previous ones on $t-1, t-2,...$ .
In online (or sequential) learning, the regret is measured by the cumulative loss it suffers along its run on a sequence of examples. We could see it as the excess loss for not consistently predicting with the optimal model.
$$
R_T = \frac{1}{T}\sum_{t=1}^T \ell(f_t(x_t), y_t) -\inf_{f\in\mathcal{F}} \left\lbrace \frac{1}{T}\sum_{t=1}^T \ell(f(x_t), y_t)\right\rbrace
$$
where the first term is the estimation error between the target and the prediction, and the second is the approximation error.
Bandits and Reinforcement Learning deal with maximizing a reward, instead of minimizing a loss. Thus, we can re-write regret as the difference between the reward that could have been achieved and what was actually achieved according to a sequence of actions,
$$
R_T = \max_a\left\lbrace \frac{1}{T}\sum_{t=1}^T r(a) \right\rbrace- \frac{1}{T}\sum_{t=1}^T r(a_t) \nonumber
$$

Thus, minimizing a loss or maximizing a reward is the same optimization problem as minimizing the regret, as defined in \cite{robbins1952}.

For instance, in the ordinary least squares regression, $\mathcal{A}=\mathcal{Y}=\mathbb{R}$, and we use the squared loss: $\ell:(a,y)\mapsto (a-y)^2$. In that case, the mean squared risk is $\mathcal{R}(f)=\mathbb{E}\left[(f(X)-Y)^2\right]$ while the target is $f^*(X)=\mathbb{E}\left[Y|X\right]$.
In the case of classification, where $y$ is a variable in $K$ categories, $\mathcal{A}$ can be a selection of a class, so $\mathcal{A}=\mathcal{Y}=\{1,\dots, K\}$. The classical loss in that case is the missclassification dummy loss $\ell(a,y)=1_{a\neq y}$, and the associated risk is the misspecification probability, $\mathcal{R}(f)=\mathbb{E}\left[1_{f(X)\neq Y}\right] = \mathbb{P}(f(X)\neq Y)$, while the target: is $f^*(X)=\displaystyle{\underset{1\leq k\leq K}{\text{argmax}}\lbrace \mathbb{P}(Y=k|X)\rbrace}$.

To go further, \cite{Mullainathan}, \cite{CharpentierFlachaireLy} or \cite{AtheyImbens2019} recently discussed connections between econometrics and machine learning, and possible applications of machine learning techniques in econometrics.

\subsection{Online learning}\label{sec:online}
In classical (or batch) learning described previously, we want to build an estimator $\widehat{f}$ from $\mathcal{D}_n=\{(x_i,y_i)\}$ such as the regret $\mathbb{E}[\mathcal{R}(\widehat{f})]-\inf_{f\in\mathcal{F}}\{R(f)\}$ is as small as possible. However, in the online learning framework, we get the data through a sequential process and the training set is changing at each iteration. Here, observations are not i.i.d, and not necessarily random.

%\subsubsection{Presentations}

Following \cite{bottou2}, assume that data become available at a sequential order, and the goal is to update our previous predictor with the new observation. To emphasize the dynamic procedure, let $t$ denote the number of available observation (instead of $n$, in order to emphasize the sequential aspect of the problem). Formally, from our sample $\mathcal{D}_t=\{(y_1,x_1),\cdots, (y_t,x_t)\}$ we can derive a model $f(x|\mathcal{D}_t)$, denoted $f_t$. The goal in online learning is to compute an update $f_{t+1}$ of $f_t$ using the new observation  $(y_{t+1},x_{t+1})$.

At step $t$, the learner gets $x_t\in\mathcal{X}$ and predicts $\widehat{y_t}\in\mathcal{Y}$, exploiting past information $\mathcal{D}_{t-1}$. Then, the real observation $y_t$ is revealed and generates a loss $\ell(\widehat{y}_t,y_t)$. Thus, $\widehat{y}_t$ is a function of $\left(x_t,(x_i,y_i)_{i=1\dots t-1}\right)$.
%In this context, despite that data is sequential, the model keeps {\em full feedback} during the training.

% At step $t$
% \begin{itemize}
%     \item the learner gets $X_t\in\mathcal{X}$
%     \item the learner predicts $\widehat{y_t}\in\mathcal{Y}$
%     \item the real prediction $y_t$ is shown to the learner
%     \item the learner suffers a loss $\ell(\widehat{y}_t,y_t)$ and updates its model
% \end{itemize}
% Thus, $\widehat{y}_t$ is a function of $\left(x_t,(x_i,y_i)_{i=1\dots t-1}\right)$ 

%\subsubsection{Examples}

%\textbf{Prediction with expert advice - aggregation}\label{agg}\\

Consider the case of forecasting with expert advice: expert aggregation.
Here, $K$ models can be used, in a supervised context, on the same objective variable $y$, $\widehat{f}_1(x|\mathcal{D}_t),\dots,\widehat{f}_K(x|\mathcal{D}_t)$. Quite naturally, it is possible a linear combination (or a weighted average) of those models,
$$\widehat{f}_{t,\omega_t}(x)=\sum_{k=1}^K \omega_{k,t}\widehat{f}_k(x|\mathcal{D}_t)$$
A natural question is the optimal choice of the weights $\omega_{k,t}$.

Assume here, as before, a sequential model. We want to predict element by element a sequence of observations $y_1,\dots ,y_T$.
At each step $t$, $K$ experts provide their forecasts $\widehat{y}_{1,t},\dots ,\widehat{y}_{K,t}$ for the next outcome $y_t$. The aggregation weights expert's prediction $\widehat{y}_{k,t}$ according to a rule in order to build its own forecast $\widehat{y}_t$
$$
\widehat{y}_t = \sum_{k=1}^K \omega_{k,t} \widehat{y}_{k,t}
$$
The weighting process is online: each instant $t$, the rule adapts the weights to the past observations and the accuracy of their respective experts, measured by the loss function for each expert $\ell(y_t, \widehat{y}_{k,t})$.

Here, the oracle (or target) is the optimal expert aggregation rule. The prediction $\widehat{y}^*$ use best possible weight combination by minimizing the loss.
The empirical regret of the aggregation rule $f$ is defined by:
$$
R_T =\frac{1}{T} \sum_{t=1}^T \ell(\widehat{y}^*_t, y_t) - \inf_{\omega \in \Omega}\left\lbrace\frac{1}{T}\sum_{t=1}^T \ell(\widehat{y}_t, y_t)\right\rbrace 
$$
where the first term is the estimation error between the target and the prediction, and the second is the approximation error.

There exist several rules for aggregation, the most popular one is probably the Bernstein Online Aggregator (BOA), described in Algorithm \ref{algo:BOA}, which is optimal with bounded iid setting for the mean squared loss.

\ 

\begin{algorithm}[H]
 \KwData{learning rate $\gamma$}
 \KwResult{Sequence $\boldsymbol{\omega}_1,\dots,\boldsymbol{\omega}_n$}
 initialization:  $\boldsymbol{\omega}_0 \leftarrow$ initial weights (e.g. $1/k$)\;
  \For{$t\in\{1,2,\dots,n\}$}{
  $L_{j,t}\leftarrow \ell(y_t,f_j(\boldsymbol{x_t}|\mathcal{D}_{t-1}))-\ell(y_t,\widehat{f}_{t-1,
  \omega_{t-1}}(\boldsymbol{x_t}))$
    $\pi_{j,t}\leftarrow \displaystyle{\frac{\pi_{j,t-1}\exp\big[-\gamma L_{j,t}(1+\gamma L_{j,t})\big]}{\exp\big[-\gamma\big]}}$ \;
    }
\caption{Bernstein Online Aggregator (BOA).}\label{algo:BOA}
\end{algorithm}

\ 

This technique, also called {\em ensemble prediction}, based on aggregation of predictive models, gives an easy way to improve forecasting by using expert forecasts directly.
{In the context of energy markets, \cite{ONeil2010} shows that a model based on aggregation of simple ones can reduce residential energy cost and smooths energy usage.}
{\cite{Levina2009} considered the case where a supplier predicts consumer demand by applying an aggregating algorithm to a pool of online predictors.}

\subsection{Bandits}\label{sec:bandits}
A related problem is the one where an agent have to choose, repeatedly, among various options but with incomplete information. Multi-armed bandits come from one-armed bandit, understand slot machines, used in casinos. Imagine an agent playing with several one-armed bandit machines, each one having a different (unknown) probability of reward associated with. The game is seen as a sequence of single arm pull action and the goal is to maximize its cumulative reward. What could be the optimal strategy to get the highest return?

In order to solve this problem and find the best empirical strategy, the agent has to explore the environment to figure out which arm gives the best reward, but at the same time must choose most of the time the empirical optimal one. It is the exploration-exploitation trade-off: each step either searching for new actions or exploiting the current best one.

{The one-armed bandit problem was used in economics in \cite{ROTHSCHILD1974185}, when trying to model the strategy of a single firm facing a market with unknown demand. In an extension, \cite{KellerRady} consider the problem of the monopolistic firm facing an unknown demand that is subject to random changes over time. Note that the case of several firms experimenting independently in the same
market was addressed in \cite{MCLENNAN1984331}}.
{The choice between various research projects often takes the form of a bandit problem. In \cite{Weitzman1979}, each arm represents a distinct research project with a random reward associated with it. The issue is to characterize the optimal
sequencing over time in which the projects should be undertaken. It shows that as novel projects provide an option value to the research, the optimal sequence is not necessarily the sequence of decreasing expected rewards. More recently, \cite{BERGEMANN1998703} and \cite{BergemannHege2005} model venture, or innovation, as a Poisson bandit model with variable learning intensity.}

%Another classical example is the choice of restaurants in a city: exploring new restaurants is the only way to discover possibly better ones, instead of relying only on our past experience, and on the long-run, this an better strategy (even if possibly, some new restaurants will be less good that previous we tried)
 
Multi-armed bandit problems are a particular case of reinforcement learning problems. However, in the bandits case the action does not impact the agent state. Bandits are an subset of model in online learning; and benefits of theoretical results under strong assumptions, most of the time to strong for real-world problems.
{The multi-armed bandit problem, originally
described by \cite{robbins1952}, is a statistical decision
model of an agent trying to optimize his
decisions while improving his information at the
same time}. {The multi-armed bandit problem and many variations are presented in detail in \cite{Git89} and \cite{berry+firstedt}. An alternative proof of the main theorem, based on dynamic programming can be found in \cite{Whi83a}. The
basic idea is to find for every arm a retirement value, and then to choose in every period the
arm with the highest retirement value.}

%\subsubsection{Concepts}

In bandits, the information that the learner gets is more restraint than in general online learning: the learner has only access to the cost (loss or reward).
At each step $t$, the learner choose $\widehat{y}_t\in\{1,\dots,K\}$. Then the loss vector $(\ell_t(1),\dots,\ell_t(K))$ is established. Eventually, the learner has access to $\ell_t(\widehat{y}_t)$.

%The classical Multi-Armed Bandit setup can be describe as a tuple $<\mathcal{A},\mathcal{R}>$
Such a problem is called $|\mathcal{A}|-${\em multi-armed bandit} in the literature,  where $\mathcal{A}$ is the set of action. The learner has $K$ arms, i.e $K$ probability distributions $(\nu_1,\dots,\nu_K)$. Each step $t$, the agent pulls an arm $a_t\in{1,\dots,K}$ and receives a reward $r_t$ following the probability distribution $\nu_{a_t}$. Let $\mu_k$ be the mean reward of distribution $\nu_k$. The value of an action $a_t$ is the expected reward $Q(a_t) = \mathbb{E}[r_t|a_t]$: if action $a_t$ at $t$ is referring to picking the $k$-th arm of the slot machine, then $Q(a_t)= \mu_k$. % As $r_t \sim \nu_{a_t}$, $r_t$ returns a reward with probability $Q(a_t)$.
The goal is to maximize the cumulative rewards $\sum_{t=1}^T r_t$. The bandit algorithm is thus a sequential sampling strategy: $a_{t+1} = f_t(a_t, r_t, \dots, a_1, r_1)$. 

To measure the bandit algorithm performance, we use the previous defined regret. Maximizing the cumulative reward becomes maximizing the potential regret, i.e. the loss of not choosing the optimal actions.\\
%For each arm $\nu_k$, we name $\mu_k$ the mean of arm k $\mu_k =\mathbb{E}_{\mathcal{R}(a_t)\sim\nu_k}[\mathcal{R}(a_t)]$. 
We note $\displaystyle{ \mu^* = \max_{a\in\{1,\dots,K\}} \lbrace\mu_a\rbrace}$ and the optimal policy is
$$
a^* =\underset{a\in\{1,\dots,K\}}{\text{argmax}}\big\lbrace \mu_a\big\rbrace
    =\underset{a\in\{1,\dots,K\}}{\text{argmax}}\big\lbrace Q(a)\big\rbrace.
$$
The regret of a bandit algorithm is thus:
$$
R_\nu(\mathcal{A}, T) = T\mu^* - \mathbb{E}\left[\sum_{t=1}^T r_t\right]
                      = T\mu^* - \mathbb{E}\left[\sum_{t=1}^T Q(a_t)\right]
$$
where the first term is the sum of rewards of the oracle strategy which always selects $a^*$, and the second is the cumulative reward of the agent's strategy.

What could be an optimal strategy ? To get a small regret, a strategy should not select to much sub-optimality arms, i.e. $\mu^* -\mu_a >0$, which requires to try all arms to estimate the values of these gaps. This leads to the exploration exploitation trade-off previously mentioned.
Betting on the current best arm $a_t = \text{argmax } \{\mu_{a_t}\}$ is called {\em exploitation}, while checking that no other arm are better $a_t \neq \text{argmax } \{\mu_{a_t}\}$ to find a lower gap is called {\em exploration}. This will be called a {\em greedy} action, since it might also be interesting to {\em explore} by selecting a non-optimal action that might improve our estimation.

%\textbf{Incremental implementation:}\\
For essentially computational reason (mainly keeping record of all the rewards on the period), it is preferred to write the value function in an incremental expression, as described in \cite{Sutton98}, %\arthur{pas de référence "Suton"}
$$
Q_{t+1} = \frac{1}{t}\sum_{i=1}^t r_i= \frac{1}{t}\left((t-1)Q_t + r_t\right)= Q_t + \frac{1}{t}(r_t - Q_t)
$$
This leads to the general update rule:
$$
\text{NewEstimate = OldEstimate + StepSize (Target - OldEstimate)},
$$
where Target is a noisy estimate of the true target, and StepSize may depends on t and a.
This value function expression, which also identifies to a gradient descent, has already be observed in \label{agg} concerning expert aggregation and will be studied again in the following.

{Recently, \cite{Misra2019} consider the case where sellers must decide, on real-time, prices for a large number of item, with incomplete demand information. Using experiments, the seller learns about the demand curve and the profit-maximizing price. The multi-armed bandit  algorithms provides an automated pricing policy, using a scalable distribution-free algorithm}.

\subsection{Reinforcement Learning: a short description}\label{sec:reinforcement}
%With the notations of supervised learning, assume now that $\boldsymbol{x}$ is some game state (say in chess, poker, go) and $y$ is optimal move. \citet{poker2011} considered a simplified version of no limit Texas Hold’em Poker: given the cards in hand, $\boldsymbol{x}$, $y$ denotes the optimal strategy. xxxx 
In the context of prediction and games (tic-tac-toe, chess, go, or video games), choosing the `best' move is complicated. Creating datasets used in the previous approaches (possibly using random simulation) is too costly, since ideally we would like to get all possible actions (positions on the chess board or hands of cards). 
As explained in \citet[page 105]{Goodfellow-et-al-2016}, ``{\em some machine learning algorithms do not just experience a fixed dataset. For example, reinforcement learning algorithms interact with an environment, so there is a feedback loop between the learning system and its experiences}''. 

%%%%%%%%%%%%%%%%%%%%%%%%%%%%%%%%%%%%%%%%%%%%%%%%%%%%%%%%%%%%%%%
\subsubsection{The concepts}\label{sec:learning:3:1}

%\carl{\textbf{plan: peu de maths}\\3phrases: env, agent, reward\\ schema\\ optimal strat and real world known pb and ref\\key concepts (a,s,r, policy, transition)\\ value fucntion\\ \textbf{puis dans 3:}\\mdp, optimal policy, bellman\\explor/exploit, model based, model free}

In Reinforcement Learning, as in Multi-armed Bandits, data is available at sequential order. But the actions depends on the environment, thus an action at a certain state could give a different reward re-visiting the same state. More specifically, at time $t$
\begin{itemize}
    \item[-] the learner takes an {\em action} $a_t\in\mathcal{A}$
    \item[-] the learner obtains a (short-term) {\em reward} $r_t\in\mathcal{R}$
    \item[-] then the {\em state} of the world becomes $s_{t+1}\in\mathcal{S}$
\end{itemize}

The states $\mathcal{S}$ refer to the different situations the agent might be in. In the maze, the location of the rat is a state of the world. The actions $\mathcal{A}$ refer to the set of options available to the agent at some point in time, across all states of the world, and therefore, actions might depend on the state. If the rat is facing a wall, in a dead-end, the only possible action is usually to turn back, while, at some crossroad, the rat can choose various actions. The rewards set $\mathcal{R}$ refer to how rewards (and possibly punishments) are distributed. It can be deterministic, or probabilistic, so in many cases, agents will compute expected values of rewards, conditional on states and actions. These notations were settled in \citet{Sutton98}, where the goal is to maximize rewards, while previously, \citet{Bertsekas96} suggested to minimze costs, with some cost-to-go functions.

As in Bandits, the interaction between the environment and the agent involves a trajectory (called also episode). The trajectory is characterized by a sequence of states, actions and rewards. The initial state leads to the first action which gives a reward; then the model is fed by a new state followed by another action and so on.

To determine the dynamics of the environment, and thus the interaction with the agent, the model relies on {transition} probabilities. It will be based on past states, and past actions, too. Nevertheless, with the Markov assumption, we will assume that transition probabilities depend only on the current state and action, and not the full history.

Let $T$ be a transition function $\mathcal{S}\times\mathcal{A}\times\mathcal{S}\rightarrow [0,1]$ where:
$$
\mathbb{P}\big[s_{t+1}=s'\big\vert s_t=s,a_t=a, a_{t-1},a_{t-2},\dots\big]=T(s,a,s').
$$
As a consequence, when selecting an action $a$, the probability distribution over the next states is the same as the last time we tried this action in the same state.

A {\em policy} is an action, decided at some state of the world. Formally policies are mapping from $\mathcal{S}$ into $\mathcal{A}$, in the sense that $\pi(s)\in\mathcal{A}$ is an action chosen in state $s\in\mathcal{S}$. Note that stochastic policies can be considered, and in that case, $\pi$ is a $\mathcal{S}\times\mathcal{A}\rightarrow [0,1]$ function, such that $\pi(a,s)$ is interpreted as the probability to chose action $a\in\mathcal{A}$ in state $s\in\mathcal{S}$. The set of policies is denoted $\Pi$.

After time step $t$, the agent receives a reward $r_t$. The goal is to maximize its cumulative reward in the long run, thus to maximize the expected return. Resuming \cite{Sutton98}, we can defined the return as the sum of the reward:
$$
G_t = \sum_{k=t+1}^T r_k
$$
Unlike in bandits approaches, here the cumulative reward is computed starting from $t$.
Sometimes the agents can receive running reward, associated to tasks  where there is no notion of final time step, so we introduce the discounted return: 
$$
G_t = \sum_{k=0}^\infty \gamma^k r_{t+1+k}
$$
where $0\leq \gamma \leq 1$ is the discount factor which gives more importance to recent reward (and can allow $G_t$ to exist).
We can also re-write $G_t$ in a recursive (or incremental way too) since $G_t=r_{t+1} + \gamma G_{t+1}$.%\carl{a developper}

To quantify the performance of an action, we introduce, as in the previous section, the action-function, or $Q$-value on $\mathcal{S}\times\mathcal{A}$:
\begin{equation}\label{eq:Q:pi}
Q^{\pi}(s_t,a_t)=\mathbb{E}_{\mathbb{P}}\left[G_t\Big\vert s_t,a_t,\pi \right]
%=\mathbb{E}_{\mathbb{P}}\left[\sum_{k\in\mathbb{N}} \gamma^k r_{t+k}\Big\vert s_t,a_t,\pi \right]
\end{equation}

In order to maximize the reward, as in bandits, the optimal strategy is characterized by the optimal policies 
$$
\pi^\star(s_t) = \underset{a\in\mathcal{A}}{\text{argmax}}\big\lbrace
Q^{\star}(s_t,a)
\big\rbrace.
$$
That function can be used to derive an optimal policy, 
and the optimal value function producing the best possible return (in sense of regret):$$
Q^{\star}(s_t,a_t) = \max_{\pi\in\Pi}\big\lbrace
Q^{\pi}(s_t,a_t)
\big\rbrace.
$$

Considering optimal strategy and regret leads to the previously mentioned exploration exploitation trade-off. As seen in the bandits section, the learner try various actions to explore the unknown environment in order to learn the transition function $T$ and the reward $R$. The exploration is commonly implemented by $\varepsilon$-greedy algorithm (described in the {bandits} section), as in Monte-Carlo methods or $Q$-learning.

{\cite{Bergemann96} provided a nice economic application of the ex\-plo\-ra\-tion-exploitation dilemma. In this model, the true value of each seller's product to the buyer is initially unknown, but additional information can be gained by experimentation. When assuming that prices are given exogeneously, the buyer's problem is a standard multi-armed bandit problem. The paper in nevertheless original since the cost of experimentation is here endogenized.}

%%%%%%%%%%%%%%%%%%%%%%%%%%%%%%%%%%%%%%%%%%%%%%%%%%%%%%%%%%%%%%%%%%%
\subsubsection{An inventory illustration}\label{sec:learning:3:2}

A classical application of such framework is the control of inventory, with limited size, when the demand is uncertain. Action $a_t\in\mathcal{A}$ denote the number of ordered items arriving on the morning of day $t$. The cost is $\underline{p}a_t$ if the individual price of items is $\underline{p}$ (but some fixed costs to order items can also be considered). Here $\mathcal{A}=\lbrace 0,1,2,\dots, m\rbrace$ where $m$ is the maximum size of storage. States $s_t=\mathcal{S}$ are the number of items available at the end of the day (before ordering new items for the next day). Here also, $\mathcal{S}=\lbrace 0,1,2,\dots, m\rbrace$. Then, the state dynamics are
$$
s_{t+1} = \big(\min\{(s_t+a_t),m\}-\varepsilon_t\big)_+
$$
where $\varepsilon_t$ is the unpredictable demand, independent and identically distributed variables, taking values in $\mathcal{S}$. Clearly, $(s_{t})$ is a Markov chain, that can be described by its transition function $T$,
$$
T(s,a,s')=\mathbb{P}\big[s_{t+1}=s'\big\vert s_t=s,a_t=a\big]=\mathbb{P}\big[\varepsilon_t = \big(\min\{(s+a),m\}-s'\big)_+\big]
$$
The reward function $R$ is such that, on day $t$, revenue made is
$$
r_{t} = - \underline{p}a_t +\overline{p}\varepsilon_t = - \underline{p}a_t+\overline{p}\big(\min\{(s_t+a_t),m\}-s_{t+1}\big)_+=R(s_t,a_t,s_{t+1})
$$
where $\overline{p}$ is the price when items are sold to consumers (and $\underline{p}$ is the price when items are purchased). Note that in order to have a more interesting (and realistic) model, we should introduce fixed costs to order items, as costs to store item. In that case
$$
r_{t} =  - \underline{p}a_t +\overline{p}\big(\min\{(s_t+a_t),m\}-s_{t+1}\big)_+- k_1 1_{a_t>0}- k_2 s_t,
$$
for some costs $k_1$ and $k_2$. Thus, reinforcement learning will appear quite naturally in economic problems, and as we will see in the next section, several algorithms can be used to solve such problems, especially when some quantities are unknown, and can only be estimated... assuming that enough observations can be collected to do so.

%%%%%%%%%%%%%%%%%%%%%%%%%%%%%%%%%%%%%%%%%%%%%%%%%%%%
%\newpage
\section{Reinforcement Learning}\label{sec:desc}
Now that most of essential notions have been defined and explained, we can focus on Reinforcement Learning principles, and possible extensions. This section deals with the most common approaches, its links with ordinary economy or finance problems and, eventually, some know difficulties of those models. 

\subsection{Mathematical context}\label{sec:maths}
Classically, a Markov property is assumed on the reward and the observations. A Markov decision process (MDP) is a collection $(\mathcal{S},\mathcal{A},T,r,\gamma)$ where $\mathcal{S}$ is a state space, $\mathcal{A}$ is an action space, $T$ the transition function $\mathcal{S}\times\mathcal{A}\times\mathcal{S}\rightarrow [0,1]$, $R$ is a reward function $\mathcal{S}\times\mathcal{A}\times\mathcal{S}\rightarrow \mathbb{R}_+$ and $\gamma\in[0,1)$ is some discount factor. A policy $\pi\in\Pi$ is a mapping from $\mathcal{S}$ to $\mathcal{A}$.

\begin{algorithm}[!ht]
 \KwData{transition function $T$ and policy $\pi$}
 \KwResult{Sequence $(a_t,s_t)$}
 initialization:  $s_1 \leftarrow$ initial state\;
  \For{$t\in\{1,2,\dots\}$}{
    $a_t \leftarrow \pi(s_t)\in\mathcal{A}$ \;
    $s_{t+1}\leftarrow T(s_t,a_t,\cdot)=\mathbb{P}\big[s_{t+1}=\cdot\big\vert s_t,a_t, \big]\in\mathcal{S}$ \;
    }
 \caption{Policy generation}
\end{algorithm}

Given a policy $\pi$, its expected reward, starting from state $s\in\mathcal{S}$, at time $t$, is
\begin{equation}\label{eq:V:pi}
V^{\pi}(s_t)=\mathbb{E}_{\mathbb{P}}\left(\sum_{k\in\mathbb{N}} \gamma^k r_{t+k}\Big\vert s_t,\pi \right)
\end{equation}
called {\em value of a state $s$ under policy $\pi$}, where $r_t=\mathbb{E}_{a}[R(s_t,a,s_{t+1})]$ when $a\sim\pi(s_t,\cdot)$ and $\mathbb{P}$ is such that $\mathbb{P}(S_{t+1}=s_{t+1}|s_t,a_t)=T(s_t,a,s_{t+1})$. Since the goal in most problem is to find a best policy -- that is the policy that receives the most reward -- define
$$
V^{\star}(s_t) = \max_{\pi\in\Pi}\big\lbrace
V^{\pi}(s_t)
\big\rbrace
$$

\begin{algorithm}[!ht]
 \KwData{policy $\pi$, threshold $\epsilon>0$, reward $R(s,a,s')$, $\forall s,a,s'$}
 \KwResult{Value of policy $\pi$, $V^\pi$}
 initialization:  $V(s)$ for all $s\in\mathcal{S}$ and $\Delta=2\epsilon$\;
    \While{$\Delta>\epsilon$}{
    $\Delta\leftarrow 0$
  \For{$s\in\mathcal{S}$}{
    $v \leftarrow V(s)$ \;
    $V(s)\leftarrow \displaystyle{\sum_{a\in\mathcal{A}} \pi(a,s)\sum_{s'\in\mathcal{s}} T(s,a,s')\big[R(s,a,s')+\gamma V(s')\big]}$ \;
    $\Delta\leftarrow \max\{\Delta,|v-V(s)|\}$
    }}
 \caption{Policy valuation}
\end{algorithm}

As in \citet{watkins_q-learning_1992}, one can define the $Q$-value on $\mathcal{S}\times\mathcal{A}$ as
$$
Q^{\pi}(s_t,a_t)=\mathbb{E}_{\mathbb{P}}\left(\sum_{k\in\mathbb{N}} \gamma^k r_{t+k}\Big\vert s_t,a_t,\pi \right)
$$
which can be written, from Bellman's equation (see \citet{bellman_1957_dynamic})
\begin{equation}\label{eq:bellman:q}
Q^{\pi}(s_t,a_t)=\sum_{s'\in\mathcal{S}} \big[r(s_t,a_t,s')+\gamma Q^{\pi}(s',\pi(s')) \big] T(s_t,a_t,s')
\end{equation}
and as previously, let 
$$
Q^{\star}(s_t,a_t) = \max_{\pi\in\Pi}\big\lbrace
Q^{\pi}(s_t,a_t)
\big\rbrace.
$$
Observe that $Q^{\pi}(s_t,a_t)$ identifies to the value function in state $s_t$ when playing action $a_t$ at time $t$ and then acting optimally. Hence, knowing the $Q$-function directly provides the derivation of an optimal policy
$$
\pi^\star(s_t) = \underset{a\in\mathcal{A}}{\text{argmax}}\big\lbrace
Q^{\star}(s_t,a)
\big\rbrace.
$$
This optimal policy $\pi^\star$ assigns to each states $s$ the highest-valued action. In most applications, solving a problem boils down to computing the optimal policy $\pi^\star$.

Note that with finite size spaces $\mathcal{S}$ and $\mathcal{A}$, we can use a vector form for $Q^{\pi}(s,a)$'s, $\boldsymbol{Q}^\pi$, which is a vector of size $|\mathcal{S}||\mathcal{A}|$. In that case, Equation (\ref{eq:bellman:q}) can be written
\begin{equation}\label{eq:bellman:q:matrix}
\boldsymbol{Q}^{\pi}=\boldsymbol{R}+\gamma \boldsymbol{P}\boldsymbol{\Pi}\boldsymbol{Q}^\pi
\end{equation}
where $\boldsymbol{R}$ is such that
$$
\boldsymbol{R}_{(s,a)} = \sum_{s'\in\mathcal{S}} r(s_t,a_t,s') T(s_t,a_t,s')
$$
and $\boldsymbol{P\Pi}$ is the matrix of size $|\mathcal{S}||\mathcal{A}|\times|\mathcal{S}||\mathcal{A}|$ that constraints transition probabilities, from $(s,a)$ to $(s',\pi(s'))$ (and therefore depends on policy $\pi$).

If we use notations introduced in section \ref{sec:reinforcement}, we have to estimate $Q(s,a)$ for all states $s$ and actions $a$, or function $V(s)$.
Bellman equation on $Q^\pi$ means that $V^\pi$ satisfies 
\begin{equation}\label{eq:V:pi:hjb}
V^{\pi}(s_t)=\sum_{s'\in\mathcal{S}} \big[r(s_t,\pi(s_t),s')+\gamma V^{\pi}(s') \big] T(s_t,\pi(s_t),s').
\end{equation}

\begin{algorithm}[H]
 \KwData{A threshold $\epsilon$, reward $R(s,a,s')$, $\forall s,a,s'$}
 \KwResult{Optimal policy $\pi^\star$}
 initialization:  $V(s)$ for all $s\in\mathcal{S}$ and $\Delta=2\epsilon$\;
    \While{$\Delta>\epsilon$}{
    $\Delta\leftarrow 0$
  \For{$s\in\mathcal{S}$}{
    $v \leftarrow V(s)$ \;
    $V(s)\leftarrow \displaystyle{\max_{a\in\mathcal{A}} \left\lbrace\sum_{s'\in\mathcal{S}} T(s,a,s')\big[R(s,a,s')+\gamma V(s')\big]\right\rbrace}$ \;
    $\Delta\leftarrow \max\{\Delta,|v-V(s)|\}$\;
    }}
\For{$s\in\mathcal{S}$}{ 
 $\pi(s)\leftarrow \displaystyle{\underset{a\in\mathcal{A}}{\text{argmax}}\left\lbrace \sum_{s'\in\mathcal{S}} T(s,a,s')\big[R(s)+\gamma V(s')\big] \right\rbrace}$\;
}
 \caption{Direct policy search}
\end{algorithm}

\

Unfortunately, in many applications, agents have no prior knowledge of reward function $r$, or transition function $T$ (but do know that it satisfies the Markov property). Thus, the agent will have to {\em explore} -- or perform actions -- that will give some feedback, that can be used, or {\em exploited}.

% explor/exploit
As discussed previously, $Q$ function is updated using
$$
Q(s,a) \leftarrow (1-\alpha) Q(s,a) + \alpha \big(r(s,a,s')+\gamma \max_{a'\in\mathcal{A}}\big\lbrace Q(s',a')\big\rbrace\big).
$$

A standard procedure for exploration is the $\epsilon$-greedy policy, mentioned already in the bandit context, where the learner makes the best action with probability $1-\epsilon$, and consider a randomly selected action with probability $\epsilon$. Alternatively, consider some exploration function that will give preference to less-visited states, using some sort of penalty
$$
Q(s,a) \leftarrow (1-\alpha) Q(s,a) + \alpha \left(r(s,a,s')+\gamma \max_{a'\in\mathcal{A}}\left\lbrace Q(s',a')+\frac{\kappa}{n_{s,a}}\right\rbrace\right).
$$
where $n_{s,a}$ denotes the number of times where state $(s,a)$ has been visited, where $\kappa$ will be related to some exploration rate. Finally, with the Boltzmann exploration strategy, probabilities are weighted with their relative $Q$-values, with 
$$
p(a) = \frac{e^{\beta Q(s,a)}}{e^{\beta Q(s,a_1)}+\dots+e^{\beta Q(s,a_n)}},
$$
for some $\beta>0$ parameter. With a low value for $\beta$, the selection strategy tends to be purely random. On the other hand, with a high value for $beta$, the algorithm selects the action with the highest $Q$-value, and thus, ceases the experiment.

\subsection{Some Dynamical Programming principles}
In Dynamic Programming, as well as in most of Reinforcement Learning problem, we use value functions to choose actions and build an optimal policy. Many algorithms of this field compute optimal policies in a fully know model in a Markov decision process environment. It is not always possible in real-world problems or too computational expensive. However, Reinforcement Learning lies on several principles of Dynamic Programming and we present here a way to obtain an optimal policy once we have found the optimal value functions which satisfy the Bellman equation: the Policy iteration.

\subsubsection{Policy iteration}

Value function $V^\pi$ satifies Equation (\ref{eq:V:pi:hjb}), or to be more specific a system of $|\mathcal{S}|$ linear equations, that can be solved % using linear programming
when all functions -- $T$ and $r$ -- are known. An alternative is to use an iterative procedure, where Bellman's Equation is seen as a updating rule, where $V_{k+1}^\pi$ is an updated version of $V_k^\pi$
\begin{equation}\label{eq:V:pi:hjb+k}
V_{k+1}^{\pi}(s_t)=\sum_{s'\in\mathcal{S}} \big[r(s_t,\pi(s_t),s')+\gamma V_k^{\pi}(s') \big] T(s_t,\pi(s_t),s').
\end{equation}
The value function $V^\pi$ is a fixed point of this recursive equation.

Once we can evaluate a policy $\pi$, \citet{howard1960mdp} suggested a simple iterative procedure to find the optimal policy, called {\em policy iteration}. The value of all action $a$ is obtained using
$$
Q^{\pi}(s_t,a)=\sum_{s'\in\mathcal{S}} \big[r(s_t,a,s')+\gamma V^{\pi}(s') \big] T(s_t,a,s'),
$$
so if $Q^{\pi}(s_t,a)$ is larger than $V^{\pi}(s_t) $ for some $a\in\mathcal{A}$, choosing $a$ instead of $\pi(s_t)$ would have a higher value. It is then possible to improve the policy by selecting that better action. Hence, a greedy policy $\pi'$ can be considered, simply by choosing the best action,
$$
\pi'(s_t) = \underset{a\in\mathcal{A}}{\text{argmax}}\lbrace Q^{\pi}(s_t,a) \rbrace.
$$ 
The algorithm suggested by  \citet{howard1960mdp} starts from a policy $\pi_0$, and then, at step $k$, given a policy $\pi_k$, compute its value $V^{\pi_k}$ then improve it with $\pi_{k+1}$, and iterate.

% \begin{algorithm}[H]% page 52
%  \KwData{function $V(s)\in\mathbb{R}$ and $\pi(s)\in\mathcal{A}$ for all $s\in\mathcal{S}$}
%  \KwResult{Optimal policy $\pi^\star$}
%  initialization\;
%  \While{$stable$ = false}{
%   $\Delta \leftarrow 0$\;
%  \While{$\Delta<\epsilon$}{
%   $\Delta \leftarrow 0$\;
%   \For{$s\in\mathcal{S}$}{
%     $v \leftarrow V^\pi(s)$ \;
%     $V(s)\leftarrow \displaystyle{\sum_{s'\in\mathcal{S}} xxx}$\;
%      $\Delta \leftarrow \max\{\Delta,|v-V(s)|\}$\;
%     }}
%   \eIf{understand}{
% \For{$s\in\mathcal{S}$}{
%  $b\leftarrow \pi(s)$\;
%  $\pi(s)\leftarrow \text{argmax}\left\lbrace xx\right\rbrace$\;
%  \eIf{$b\neq \pi(s)$}{xxx}
% }}}
% \caption{Policy iteration, \citet{howard1960mdp}}
% \end{algorithm}

Unfortunately, such a procedure can be very long, as discussed in \citet{Bertsekas96}. And it assumes that all information is available, which is not the case in many applications. As we will see in the next sections, it is then necessary to sample to learn the model -- the transition rate and the reward function.

\subsubsection{Policy Iteration using least squares}\label{sec:comp:11}
%\textbf{Example: Policy Iteration using least squares}
$Q^{\pi}(s,a)$ is essentially an unknown function, since it is the expected value of the cumulated sum of discounted future random rewards. As discussed in Section \ref{sec:online}, a natural stategy is to use a parametric model, ${Q}^{\pi}(s,a,\boldsymbol{\beta})$ that will approximate $Q^{\pi}(s,a)$. Linear predictors are obtained using a linear combination of some basis functions,
$$
Q^{\pi}(s,a,\boldsymbol{\beta}) = \sum_{j=1}^k  \psi_j(s,a)\beta_j =  \boldsymbol{\psi}(s,a)^\top\beta_j,
$$
for some simple functions $\psi_j$, such as polynomial transformations. With the notation of section \ref{sec:learning:3:1}, write $\boldsymbol{Q}^\pi=\boldsymbol{\Psi} \boldsymbol{\beta}$. Thus, substituting in equation (\ref{eq:bellman:q:matrix}), we obtain
$$
\boldsymbol{\Psi} \boldsymbol{\beta} \approx \boldsymbol{R}+\gamma\boldsymbol{P}\boldsymbol{\Pi}\boldsymbol{\Psi} \boldsymbol{\beta}\text{ or }
\big(\boldsymbol{\Phi} -\gamma\boldsymbol{P}\boldsymbol{\Pi}\boldsymbol{\Psi}\big) \boldsymbol{\beta}\approx\boldsymbol{R}.
$$
As in section \ref{sec:learning:3:1}, we have an over-constrained system of linear equations, and the least-square solution is
$$
\boldsymbol{\beta}_\star = \big((\boldsymbol{\Psi} -\gamma\boldsymbol{P}\boldsymbol{\Pi}\boldsymbol{\Psi})^{\top}(\boldsymbol{\Psi} -\gamma\boldsymbol{P}\boldsymbol{\Pi}\boldsymbol{\Psi})\big)^{-1}(\boldsymbol{\Psi} -\gamma\boldsymbol{P}\boldsymbol{\Pi}\boldsymbol{\Psi})^\top\boldsymbol{R}.
$$
This is also called {\em Bellman residual minimizing approximation}. And as proved in \citet{Nedic:2003} and \citet{Lagoudakis:2003}, for any policy $\pi$, the later can be written
$$
\boldsymbol{\beta}_\star = \big(\underbrace{\boldsymbol{\Psi}^{\top}(\boldsymbol{\Psi} -\gamma\boldsymbol{P}\boldsymbol{\Pi}\boldsymbol{\Psi})}_{=\boldsymbol{A}}\big)^{-1}\underbrace{\boldsymbol{\Psi} ^\top\boldsymbol{R}}_{=\boldsymbol{b}}.
$$
Unfortunately, when rewards and transition probability are not given, we cannot use (directly) the equations obtained above. But some approximation, based on previous $t$ observed values can be used. More precisely, at time $t$ we have a sample $\mathcal{D}_t=(s_i,a_i,r_i)$, and we can use algorithm \ref{algo:LS}.

\

\begin{algorithm}[!ht]
 \KwData{Policy $\pi$, $\gamma$, sample $\mathcal{D}_t$ and basis functions $\psi_j$}
 \KwResult{Optimal $\pi$}
 initialization $\widehat{\boldsymbol{A}}\leftarrow \boldsymbol{0}$ and $\widehat{\boldsymbol{B}}\leftarrow \boldsymbol{0}$\;
  \For{$i\in\{1,2,\cdots,t-1\}$}{
    $\widehat{\boldsymbol{A}} \leftarrow \widehat{\boldsymbol{A}} + \psi(s_i,a_i)\big(\psi(s_i,a_i)-\gamma \psi(s_{i+1},\pi(s_{i+1}))\big)^\top$ \;
    $\widehat{\boldsymbol{b}} \leftarrow \widehat{\boldsymbol{b}} + \psi(s_i,a_i)r_i$ \;
    }
$\widehat{\boldsymbol{\beta}}_\star \leftarrow \widehat{\boldsymbol{A}}^{-1}\widehat{\boldsymbol{b}} $    \;
  $\pi^\star(s) \leftarrow \displaystyle{\underset{a\in\mathcal{A}}{\text{argmax}}\left\lbrace \psi(s,a)^\top \widehat{\boldsymbol{\beta}}_\star\right\rbrace}$
 \caption{Least square policy iteration}\label{algo:LS}
\end{algorithm}

If states and actions are uniformely observed on those $t$ past values, $\widehat{\boldsymbol{A}} $ and $\widehat{\boldsymbol{b}} $ converge respectively towards ${\boldsymbol{A}} $ and ${\boldsymbol{b}} $ and therefore, $
\widehat{\boldsymbol{\beta}}_\star =\widehat{\boldsymbol{A}}^{-1}\widehat{\boldsymbol{b}}
$ is a consistent approximation of ${\boldsymbol{\beta}}_\star$.

% \begin{figure}[h!]
%   \begin{center}
%   \includegraphics[width=\textwidth]{rl_learnings.png}
%   \end{center}
%   \label{fig:rldiagrams}
%   \caption{}
%   \flushleft{\carl{Comparison of the backup diagrams of Monte-Carlo, Temporal-Difference learning, and Dynamic Programming for state value functions. (Image source: David Silver’s RL course lecture 4: “Model-Free Prediction”)}}
% \end{figure}

\subsubsection{Model-Based vs Model-Free Learning}\label{sec:model-based}
{\em Model-based} strategies are based on a fully known environment. We can learn about the state transition $T(s_t,a_t,s_{t+1})=\mathbb{P}(S_{t+1}=s_{t+1}|s_t,a_t)$ and the reward function $R(s_t)$ and find the optimal solution using dynamic programming.
Starting from $s_0$, the agent will chose randomly selection actions in $\mathcal{A}$ at each step. Let $(s_i,a_i,s_{i+1})$ denote the simulated set of present state, present action and future state. After $n$ generations, the empirical transition is
$$
\widehat{T}_n(s,a,s') = \frac{\sum_i \boldsymbol{1}_{(s,a,s')}(s_i,a_i,s_{i+1})}{\sum_i \boldsymbol{1}_{(s,a)}(s_i,a_i)}
$$
and
$$
\widehat{R}_n(s,a,s') = \frac{\sum_i R(s_i,a_i,s_{i+1})}{\sum_i \boldsymbol{1}_{(s,a,s')}(s_i,a_i,s_{i+1})}
$$
By the law of large numbers, $\widehat{T}_n$ and $\widehat{R}_n$ will respectively converge towards $T$ and $R$, as $n$ goes to infinity. This is the {\em exploration} part.

That strategy is opposed to so-called {\em model-free} approaches. 

In the next sections, we will describe classical model-free algorithms: Temporal-Difference (TD), Policy Gradient and Actor-Critic. 
For the first one, we will focus on one significant breakthroughs in reinforcement learning, the $Q$-learning (introduced in \cite{Watkins89}), an off-policy TD control model. As TD approach, it will necessitate to interact with the environment, meaning that it will be necessary to simulate the policy, and to generate samples, as in the generalized policy iteration (GPI) principle, introduced in \citet{Sutton98}. Recent works using neural network, like Deep Q-Network (DQN) show impressive results in complex environment.\\

%\carl{In the following, we detail {\em model-free} approaches (Value based, Policy based or both), as in economy or finance problems we deal with incomplete information. - pas sûr}

% \begin{algorithm}[H]
%  \KwData{A threshold $\epsilon$, ... $n\in\mathbb{N}$}
%  \KwResult{Optimal policy $\pi^\star$}
%  \For{$i\in\{1,2,\dots,n\}$}{
%  ...
%     }
% \For{$s\in\mathcal{S}$}{ 
%  $\pi(s)\leftarrow \underset{a\in\mathcal{A}}{\text{argmax}}\left\lbrace \sum_{s'\in\mathcal{S}} T(s,a,s')\big[R(s,a,s')+\gamma V(s')\big] \right\rbrace$\;
% }
%  \caption{Direct policy search, Monte Carlo}
% \end{algorithm}

\subsection{Some Solution Methods}
%\carl{manque de ref, manque de details}
Here is presented briefly some common methods to solve Reinforcement Learning problems.

\subsubsection{\textit{Q}-learning}\label{value}
$Q$-learning was introduced in \citet{watkins_q-learning_1992}. Bellman Equation (\ref{eq:bellman:q}) was
$$
Q^{\pi}(s_t,a_t)=\sum_{s'\in\mathcal{S}} \big[R(s_t,a_t,s')+\gamma Q^{\pi}(s',\pi(s')) \big] T(s_t,a_t,s'),
$$
and the optimal value was satisfies
$$
Q^{\star}(s_t,a_t)=\sum_{s'\in\mathcal{S}} \big[R(s_t,a_t,s')+\gamma V^{\star}(s') \big] T(s_t,a_t,s')\text{ where }
V^{\star}(s')=\max_{a'\in\mathcal{A}}\big\lbrace Q^{\star}(s',a') \big\rbrace.
$$
Thus, $Q$-learning is based on the following algorithm: starting from $Q_0(s,a)$, at step $k+1$ set
$$
Q_{k+1}(s,a)=\sum_{s'\in\mathcal{S}} \big[R(s,a,s')+\gamma \max_{a'\in\mathcal{A}}\big\lbrace Q_k(s',a') \big\rbrace \big] T(s,a,s').
$$
This approach is used in \cite{DQN} where the $Q$-function, i.e. value-function, is approximated by a neural network.

\subsubsection{Policy Optimization}\label{policy}
In order to avoid computing and comparing the expected return of different actions, as in $Q$-learning, an agent could learn directly a mapping from states to actions. 
Here, we try to infer a parameterized policy $\pi(a|s,\theta)$ that maximizes the outcomes reward from an action on an environment. Policy learning converges faster than Value-based learning process and allows continuous action space of the agent as the policy is now a parameterized function depending on $\theta$. An infinite number of actions would be computationally too expensive to optimize.
This approach is based the on Policy Gradient Theorem from \cite{Sutton98}.

\subsubsection{Approximate Solution Methods: Actor-Critic}
Actor-Critics aim to take advantage of both Value and Policy approaches . By merging them, it can benefit of continuous and stochastic environments and faster convergence of Policy learning, and sample efficiency and steady of Value one.
In the Actor-Critic approach, two model interact in order to gives the best cumulative reward. Using simultaneously an actor, which updates the policy parameter, and a critic which updates the value function or action-value function, this model is able to learn complex environments as well as complex Value-functions.

\subsection{Inverse Reinforcement Learning}\label{sec:inverse}
In the econometric literature, this problem can be found in many articles published in the 80's, such as  \citet{Miller84} in the context of job matching and occupational choice, \citet{NBERc10045} on the rate of obsolescence of patents, and research gestation lags, \citet{Wolpin} on the estimation of a dynamic stochastic model of fertility and child mortality, \citet{Pakes86} on optimal investment strategies or \citet{Rust87} on replacement of bus engines, where structural models are used to better understand human decision making. \citet{HotzMiller93}, \citet{AguirregabiriaMira02} or more recently \citet{MagnacThesmar02} or \citet{su-Judd} mentioned the computational complexity of such algorithms on economic applications.

Most of those approaches are related to the literature on dynamic discrete choice model (see \citet{AGUIRREGABIRIA201038} for a survey, or \citet{semenova2018machine} for connections with machine learning tools). In those models, there is a finite set of possible actions $\mathcal{A}$, as assumed also in the previous descriptions, and they focus on conditional choice probability, which is the probability that choosing $a\in\mathcal{A}$ is optimal in state $s\in\mathcal{S}$,
$$
\text{ccp}(a|s)=\mathbb{P}[a\text{ is optimal in state }s]=\mathbb{P}\big[\{Q(a,s)\geq Q(a',s),\forall a'\in\mathcal{A}\}\big].
$$
Assuming that rewards have a Gumbel distribution, we obtain a multinomial logit model, where the log-odds ratios are proportional to the value function. For instance in the bus-repair problem of \citet{Rust87}, the state $s$ is the mileage of the bus, and the action $a$ is in the set $\{\text{opr},\text{rep}\}$ (either operate, or replace). Per period, the utility is
$$
U_{\theta}(s_t,\varepsilon_t,a)=\varepsilon_t+u_{\theta}=\varepsilon_t+\left\lbrace
\begin{array}{ll}
  -OC_{\theta}(s_t)   &  \text{ if }a=\text{opr}\\
  -RC-OC_{\theta}(0)   &  \text{ if }a=\text{rep}\\ 
\end{array}
\right.
$$
where $RC$ is some (fixed) replacing cost, $OC_{\theta}$ is the operating cost (that might depend on some parameter ${\theta}$), and $\varepsilon_t$ is supposed to have a Gumbel distribution. The respective costs are supposed to be known

Then
$$
\text{ccp}_{\theta}(a|s)=\frac{\exp[v_{\theta}(s,a)]}{\exp[v_{\theta}(s,\text{opr})]+\exp[v_{\theta}(s,\text{rep})]}
$$
where $v_{\theta}(s,a)=u_{\theta}(s,a)+\beta EV_{\theta}(s,a)$ where $ES_{\theta}(s,a)$ is the unique solution of
$$
EV_{\theta}(s,a) = \int \log\big[u_{\theta}(s,\text{opr})+u_{\theta}(s',\text{opr})+\beta (EV_{\theta}(s,\text{opr})+ EV_{\theta}(s',\text{rep}))\big] T(s'|s,a)
$$
\citet{HotzMiller93} proved that the mapping between conditional choice probabilities and choice
specific value function is invertible.
As discussed in \citet{su-Judd}, based on observed decisions made by the superintendent of maintenance of the bus company, structural estimation is computationally complex. 

The main idea of inverse reinforcement learning (or learning from demonstration, as defined in \citet{Schaal96}) is to learn the reward function based on the agent's decisions, and then find the optimal policy (the one that maximizes this reward function) using reinforcement learning techniques. Similar techniques are related to this idea. In imitation learning (also called behavioral cloning in \citet{Bain1995AFF}), we learn the policy using supervised learning algorithms, based on the sample of observations $\{(s_i,a_i)\}$, that is unfortunately not
distributed independently and identically in the state-action space. In apprenticeship learning, we try to find a policy that perform as well as the expert policy, as introduced in \citet{abbeel04}. \citet{rothkopf2011preference} mentioned applications of reinforcement learning on preference elicitation, extended in  \citet{klein2012inverse}. See \citet{ng2000algorithms} for a survey of various algorithms used in inverse reinforcement learning, as well as  \citet{abbeel04}.

% \subsection{Mathematical context}\label{sec:maths}
% \input{reinforcement/rl_maths}

% \subsubsection{Model-Based vs Model-Free Learning}\label{sec:model-based}
% \input{reinforcement/model_based_vs_free}

% \subsection{Policy iteration}
% \input{reinforcement/policy_iter}

% \subsection{Value learning with Temporal Difference}
% \input{reinforcement/temp_diff} 
% % include TD and Q-learning

% \subsection{Policy Learning and extension}
% \input{reinforcement/policy_learning}

% \subsection{Inverse Reinforcement Learning}\label{sec:inverse}
% \input{reinforcement/inverse_rl}

%%%%%%%%%%%%%%%%%%%%%%%%%%%%%%%%%%%%%%%%%%%%%%%%%%%%
%\newpage
\section{Applications}
\subsection{Applications in Economic Modeling}\label{sec:econ}

If it is possible to find a framework very similar to the one use in reinforcement learning in old economic literature (see for instance the seminal thesis \citet{Hellwig1973}), as mentioned in \citet{arthur1991} or  \citet{BARTO199135}, two survey of reinforcement learning techniques in computational economics, published thirty years ago. Recently, \citet{Hughes14} updated the survey on applications of reinforcement learning to economic problems with up-to-date algorithms.

\subsubsection{Consumption and Income Dynamics}

Consider an infinitely living agent, with utility $u(c_t)$ when consuming $c_t\geq 0$ in period $t$. That agent receives random income $y_t$ at time $t$, and assume that $(y_t)$ is a Markov process with transition $T(s,s')=\mathbb{P}[y_{t+1}=s'|y_t=s]$. Let $w_t$ denote the wealth of the agent, at time $t$, so that $w_{t+1}=w_t+y_t-c_t$. Assume that the wealth must be non-negative, so $c_t\leq w_t+y_t$. And for convenience, $w_0=0$, as in \citet{Lettau}. At time $t$, given state $s_t=(w_t,y_t)$, we seek $c_t^\star$ solution of 
$$
v(w_t,y_t)=\max_{c\in[0,w_t+y_t]}\left\lbrace
u(c)+\gamma\sum_{y'} \big[v(w_t+y_t-c,y')\big] T(y_t,y')
\right\rbrace
$$
This is a standard recursive model, discussed in \citet{Ljungqvist2012} or \citet{HansenSargent2013}, assuming that utility function $u$ is continuous, concave, strictly increasing and bounded, the value function $v$ is itself continuous, concave, strictly increasing and bounded in wealth $w_t$, and gives a unique decision function $c^\star(w_t, y_t)$.
\citet{StokeyLucas1989} extented that model to derive a {\em general dynamic decision problem} where income $y$ is now a state $s\in\mathcal{S}=\lbrace s_1,\dots,s_n\rbrace$, and consumption $c$ is now an action $a\in\mathcal{A}=\lbrace a_1,\dots,a_m\rbrace$. Utility is now a function of $(s,a)$, and it is assume that the state process $(s_t)$ is a Markov chain, with transition matrix $\boldsymbol{T}_a$ (and transition function $T_a$). The decision problem is written as a dynamic problem
$$
v(s) = \max_{a\in\mathcal{A}}\big\lbrace u(s,a)+\gamma\mathbb{E}_{s'\sim T_a} \big[v(s')\big]
\big\rbrace
$$
Using contraction mapping theorems, there is a unique solution $v^\star$ to this problem, that can be characterized by some decision function $\pi^\star:\mathcal{S}\mapsto \mathcal{A}$ that prescribes the best action $\pi^\star(s)$ in each state $s$.
$$
v^{\pi}(s)=u(s,\pi(s))+\gamma\mathbb{E}_{s'\sim T_a} \big[v^\pi(s')\big]
$$
The solution can be obtained easily using some matrix formulation, $\boldsymbol{v}^\pi=(\mathbb{I}_n-\gamma \boldsymbol{T}^\pi)^{-1}\boldsymbol{u}^\pi$, where $\boldsymbol{v}^\pi=(v^{\pi}(s_i))\in\mathbb{R}^n$, $ \boldsymbol{T}^\pi=[T^{\pi(s_i)}(s_j)]$ is a $n\times n $ matrix, and $\boldsymbol{u}^\pi = (s_i,\pi(s_i))\in\mathbb{R}^n$. Once $v^\pi$ is obtained for any policy $\pi$, then $v^\star$ is the maximum value. \citet{StokeyLucas1989} gives several rules of thumb to solve that problem more efficiently, inspired by \citet{holland:mlaia86}. 

In the context of multiple agents, \cite{Kiyotaki-Wright} describes an economy with three indivisible goods, that could be stored, but with a cost, and three types of agents, infinitely living, favoring one of the good. In \cite{BASCI19991569},  agents do not know the equilibrium strategies and act according to some randomly held beliefs regarding the values of the possible actions. Agents have opportunities of both learning by experience, and by imitation. \cite{BASCI19991569} observes that the presence of imitation either speeds up social convergence to the theoretical Markov-Nash equilibrium or leads every agent of the same type to the same mode of suboptimal behavior. We will discuss Nash equilibrium with multiple agents in the next section.

\subsubsection{Bounded Rationality}

\citet{simon1972theories} discussed the limits of the rationality concept, central in most economic models, introducing the notion of bounded rationality, related to various concepts that were studied afterwards, such as 
bounded optimality (as in \citet{RussellDevika} with possible limited thinking time, or memory constraints) or computational rationality (as defined in \citet{Gershman273}) minimal rationality (such as \citet{Cherniak:1986} where minimal sets of conditions to have rationality are studied), ecological or environmental rationality (with a close look at the environment, that will influence decisions, as discussed in \citet{gigerenzer1996reasoning}). More recently, \citet{kahneman2011} popularized this concept with the two modes of thought: {\em system 1} is fast, instinctive and emotional while {\em System 2} is slower, more deliberative, and more logical. \citet{simon1972theories} suggests that bounded rationality can be related to uncertainty, incomplete information, and possible deviations from the original goal, emphasizing the importance of heuristics to solve complex problems, also called practical rationality (see \cite{rubinstein:98a} of \citet{AUMANN19972} for some detailed survey). 
Recently, \citet{Leimar2019} suggested that adaptive and reinforcement learning leads to bounded rationality, while \citet{Abel19} motivates reinforcement learning as a suitable formalism for studying boundedly rational agents, since ``{\em at a high level, Reinforcement Learning unifies learning and decision making into a single, general framework}''. 

\citet{simon1972theories} introduce dthe problem of {\em infinite regress}, where agents are spending more resources on finding the optimal simplification of the problem than solving the original problem. This simplification problem is related to the sparsity issue in standard supervised learning. \citet{gaba14-sparsity-based-a} discussed algorithms for finding a sparse model, either with short range memory, or focusing on {\em local thinking}, as defined in \citet{GennaioliSchleifer} (where agents combine data received from the external world with information retrieved from memory to evaluate a hypothesis). Reinforcement learning provides powerful tools to solve complex problems, where agents are suppose to have bounded rationality. And the literature (in reinforcement learning) has developed sereval measures for evaluating the capacity of an agent to effectively explore its environment. The first one is the regret of an agent, which measures how much worse the agent is relative to the optimal strategy (that could be related to unbounded rationality). The second one is the sample complexity (or computational complexity) which measures the number of samples an agent need before it can act near-optimally, with high probability.

\subsubsection{Single firm dynamics}

\citet{Jovanovic82} gave the framework for most models dealing with industry dynamics with Bayesian learning. In a model of competition between firms with multiple equilibrium, firms are engaged in an adaptive process, where they learn how
to play an equilibrium of the game, as in \citet{Fudenberg98}. In those models, firms know the model that describes the environment, but there are uncertainties. So agents will learn over time about these elements, when new information arrives. Note that this approach is different from the one in evolutionary game theory (as in \citet{Samuelson}) for instance, where agents might not even know that they play a game.

Consider a monopolistic firm, taking actions $a_t\in\mathcal{A}$ -- say investment decisions -- in order to maximize its expected discounted inter-temporal profit. States of the world are $s_t\in\mathcal{S}$, and we assume that they can be modeled via a Markov process. If future investments are uncertain, it can be assumed that the first will use the same optimal decision rule that the one it uses at time $t$, taking into account available information. Let $r_t$ denote the profit obtained at time $t$.

In economic literature, rational expectations were usually considered in early models, meaning that the expectation is computed under the true transition probability. Nevertheless, \citet{CyertDeGroot} or \citet{FELDMAN1987297} suggested that the first should learn this transition probability $\pi$, and a Bayesian framework was considered. Starting from a prior belief, transition probabilities $T$ are supposed to belong to some space $\mathcal{T}$, and experience is used to update mixing probabilities on $\mathcal{T}$. \cite{Sargent93} considered a weaker updating rule, simpler (related to linear approximations in Bayesian models) but not optimal, usually called {\em adaptative learning}. In that case, belief at time $t$, $T_t(s,a,s')$ is a weighted sum of $T_{t-1}(s,a,s')$ and some distance between $T(s,a,s')$ and $(s_{t-1},a_{t-1},s_t)$ (through some kernel function). If the weight related to the new observation is of order $1/t$, {\em recursive least squares learning} is obtained; if weights are constant, adaptative learning is here faster than standard Bayesian learning, which is usually seen as a good property when there are shocks in the economy.

\citet{erev_predicting_1998} explicitly introduced the idea of {\em stock of reinforcement}, corresponding to the standard $Q$-function. and for any action-state pair $(a,s)$, the updating rule is
$$
Q_{t+1}(a,s) \leftarrow Q_{t}(a,s) + \gamma_t k\big((a,s)-(a_t,s_t)\big)
$$
where some kernel $k$ is considered. Recently, \citet{ItoReguant} used reinforcement learning to describe sequential energy markets.

\subsubsection{Adaptative design for experiments}

Most experiments are designed to inform about the impact of choosing a policy, among various that can be considered. And more precisely, as discussed in \citet{Kasy2019AdaptiveTA}, the question {\em which program will have the largest effect} is usually preferred to the question {\em does
this program have a significant effect}, in many cases, see \citet{chattopadhyay_duflo:04} and more recently \citet{Athey_Imbens_2016}, and references therein. If dynamic experiments are considered, there are usually several waves, and the optimal experimental design would usually learn from earlier waves, and assign more experimental agents to the better-performing treatments in future waves. Thus, this policy choice problem is a finite-horizon dynamic stochastic optimization problem. \citet{Thompson} introduced this idea of adaptive treatment assignment, and \citet{weber1992} proved that this problem can be expressed using multi-armed bandits, and the optimal solution to this bandit problem is to choose the arm with the to the highest Gittins index, that can be related to the so-called Thompson sampling strategy. Thompson sampling simply assigns the next wave of agents to each treatment with frequencies proportional to the probability that that each treatment is the optimal one.

As explained in \citet{Kasy2019AdaptiveTA}, standard experimental designs are geared toward point estimation and hypothesis testing. But they consider the problem of treatment assignment in an experiment with several non-overlapping waves, where the goal is to choose among a set of possible policies (here treatments). The optimal experimental design learns from earlier waves, and assigns more experimental units to the better-performing treatments in later waves : assignment probabilities are an increasing concave function of the posterior probabilities that each treatment is optimal. They provide theoretical results to this {\em exploration sampling} design.

\subsection{Applications in Operations Research and Game Theory}\label{sec:games}
Probably more interesting is the case where there are multiple strategic agents, interacting (see \cite{zhang2019multiagent} for a nice survey). But before, let us mention the use of reinforcement learning techniques in operation research, and graphs.

\subsubsection{Traveling Salesman}

A graph $(E,V)$ is a collection of edges $E$ (possibly oriented, possibly weighted) and vertices (or nodes) $V$. There are many several classical optimization problems on graphs. In the traveling salesman problem, we want to find a subgraph $(E^\star,V)$ (with $E^\star\subset E$) which forms a cycle of minimum total weight that visits each node $V$ at least once. But one might also think of max-flow or max-cut problems, or optimal matching on bipartite graphs (see \citet{Alfred} for more examples, with economic applications). In several problems, we seek an optimal solution, which can be a subset $V^\star$ or $E^\star$, of vertices or edges. In the traveling salesman problem (TSP), given an order list of nodes $V'$ that defines a cycle ($E^\star\subset E$ is the subset of edges $\{(V'_{i},V'_{i+1})\}$ with $V\subset V$ and $V'_{i},V'_{i+1}\in E$ for all $i$), the associated loss function is
$$
\ell(V')=\sum_{i\in |V'|} \big(w(V'_{i},V'_{i+1})\big),\text{ with }V'_{|V'|+1}=V_1.
$$
Most TSP algorithms are sequential, which will make reinforcement learning perfectly appropriate here. For instance, the 2-opt algorithm (developed in \citet{flood1956tsp} and \citet{croes1958method}) suggests to iteratively remove two edges and replace these with two different edges that reconnect the fragments created by edge removal into a shorter tour (or that increases the tour least),
$$
(i^\star,j^\star) = \underset{i,j=1,\dots,|V'|}{\text{argmin}}\left\lbrace \ell(V')-\ell(\tilde{V}^{(ij)})\right\rbrace,\text{ where }\tilde{V}^{(ij)}_k=
\left\lbrace
\begin{array}{l}
V'_j\text{ if }k=i\\
V'_i\text{ if }k=j\\
V'_k\text{ otherwise}
\end{array}
\right.
$$
Other popular techniques are for instance Christophides algorithm (developed in \citet{christofides76}) or some evolutionary model inspired by ant colonies (as developed in \citet{iridia:aco}). Here also, it can be interesting to explore possibly non-optimal moves on a short term basis (in the sense that locally they end-up in a longer route)
Such sequential techniques can be formulated using the framework of reinforcement learning. The states $\mathcal{S}$ are subsets of edges $E$ in the context of TSP that form a cycle. In the 2-opt algorithm, actions $\mathcal{A}$ are nodes that will be permuted. Rewards are related to changes in the loss function (and the non-discounted sum of rewards is considered here). The nearest neighbour algorithm (which is a greedy algorithm) or cheapest insertion (as defined in \citet{Rosenkrantzetal74}) can also be seen with a reinforcement learning algorithm. States $\mathcal{S}$ are subsets of edges $E$ that form partial cycles, and the action $\mathcal{A}$ means growing the route with one node, by inserting it optimally. The rewards is related to the change in the tour length. That idea was developed in \citet{GamDor95} recently, or \citet{dai2017learning} for a recent survey of reinforcement learning techniques in the context of optimization over graphs.

\citet{deudonTSP} provides insights on how efficient machine learning algorithms could be adapted to solve combinatorial optimization problems in conjunction with existing heuristic procedures. In \citet{bello2016neural} the heuristic procedure is replaced by some neural networks. Despite the computational expense, an efficient algorithm  is obtained.

\subsubsection{Stochastic Games and Equilibrium}\label{Sec_Games}

Consider $n$ players, each of them taking actions $a_i\in\mathcal{A}_i$ and receives a reward $r_i$. Let $\boldsymbol{a}=(a_1,\dots,a_n)\in\mathcal{A}$ and $\boldsymbol{r}=(r_1,\dots,r_n)$. Note that $r_i$ is defined on $\mathcal{S}\times\mathcal{A}$. When $\mathcal{S}$ is a singleton (and there is no uncertainty), it is a simple repeated game (or matrix game). A policy $\pi_i$ maps $\mathcal{S}$ into $\mathcal{A}_i$. Let $\boldsymbol{\pi}=(\pi_1,\dots,\pi_n)$, and $\boldsymbol{\pi}_{-i}$ the collection of all component policies. Thus, $\boldsymbol{\pi}=(\pi_i,\boldsymbol{\pi}_{-i})$ means that player $i$ uses policy $\pi_i$ while competitors follow $\boldsymbol{\pi}_{-i}$.

\citet{maskin.tirole:theory1} introduced the concept of {\em Markov perfect equilibrium}, which is a set of Markovian policies $\boldsymbol{\pi}=$ which simultaneously forms a Nash equilibrium, as discussed in details in \cite{HORST200583} or \citet{ESCOBAR201392}. The existence results of such equilibrium are usually performed in two step: first, we should prove that given any policies chosen by opponents, $\boldsymbol{\pi}_{-i}$, there is a unique solution $V_i^\star(s)$; and then we prove that the static game has a Nash equilibrium for any state $s$. For the first step, the set of best response for player $i$ is $\Pi_i(\boldsymbol{\pi}_{-i})$ such that $\pi_i^\star\in\Pi_i(\boldsymbol{\pi}_{-i})$ if and only if for any $\pi_i$ and $s\in\mathcal{S}$, $V_i^{(\pi_i^\star,\boldsymbol{\pi}_{-i})}(s)\geq V_i^{(\pi_i,\boldsymbol{\pi}_{-i})}(s)$. And a Nash equilibrium is a collection of policies $\boldsymbol{\pi}=(\pi_1,\dots,\pi_n)$ such that for each player $i$, $\pi_i\in\Pi_i(\boldsymbol{\pi}_{-i})$. And therefore, no player can do better when changing policies, when other players continue to use their own strategies.

\citet{littman_markov_1994} used $Q$-learning algorithms for zero-sum stochastic games, with two players. More precisely,
$$
V_1(s) = \max_{\pi}\left\lbrace
\min_{a_2\in\mathcal{A}_2}\left\lbrace
\sum_{a_1\in\mathcal{A}_1} \pi(s,a_1) Q_1(s,\boldsymbol{a})
\right\rbrace
\right\rbrace= -V_2(s).
$$

\citet{erev_predicting_1998} proved that in many games, a one-parameter reinforcement learning model robustly outperforms the equilibrium predictions. Predictive power is improved by adding a {\em forgetting} property and valuing {\em experimentation}, with strong connections with rationality concepts. In the context of games, \cite{FRANKE2003367} applies the approach of reinforcement learning to \cite{arthur1994}'s El Farol problem, where repeatedly a population of agents decides to go to a bar or stay home, and going is enjoyable if, and only if, the bar is not crowded. 

The main difficulty arising when several agents are learning simultaneously in a game is that, for each player, the strategy of all the other players becomes part of the environment. Hence the environment dynamics do not remain stationary as the other players are learning as they play. In such context, classical single agent based reinforcement learning algorithms may not converge to a targeted Nash equilibrium, and typically cycles in between several of them, see \cite{hart2003uncoupled}. As observed by \citet{erev_predicting_1998} or in a more general setting by \cite{perolat2018actor}, stabilizing procedures such as fictitious play (\cite{robinson1951iterative}) allows to reach Nash equilibria in some (but not all,  \cite{shapley1964some}) multi Agent learning setting. \cite{48766} observed  that such property also extends to the asymptotic mean field game setting introduced by \cite{huang2006} and  \cite{LASRY2006619,LASRY2006679}, where the size of the population is infinite and shares mean field interaction. Multi-Agent reinforcement learning algorithms still lack scalability when the number of agents becomes large, a weakness that mean field games asymptotic properties may hopefully allow to partially overcome.

%As discussed in \citet{Harrington}, the diffusion and evolution of pricing algorithms raise several issues for competition policy.

\subsubsection{Auctions and real-time bidding}

The majority of online display ads are served through real-time bidding.  To place an ad automatically, and optimally, it is critical for advertisers to have a learning algorithm that cleverly bids.
\cite{Schwind07} did show that seeing the bid decision process as a reinforcement learning problem, where the state space is represented by the auction information and the campaign's real-time parameters, while an action is the bid price to set, was very promising. More recently, \cite{auction-dar-09}, \cite{auction-zhang-14}, \cite{auction-cai-17} or \cite{auction-zhao-18} use reinforcement learning algorithms to design a bidding strategy.

As pointed out by recent articles, the scalability problem from the large real-world auction volume, and campaign budget, is well handled by state value approximation using neural networks. \cite{dtting2017optimal} and \cite{Feng18} suggested to use deep reinforcement learning (with deep neural networks) for the automated design of optimal auctions. Even if the optimal mechanism is unknown, they obtain very efficient algorithm, that outperforms more classical ones.

%\citet{Shapley1095}

%\citet{Kianercy} studied convergence of games with $2$ players.

%\romu{Ajouter several players MFG}

%\cite{huang2006}, \cite{LASRY2006619} and \cite{LASRY2006679}

\subsubsection{Oligopoly and dynamic games}

As in the monopolistic case, the profit of firm $i$ will depend on its investing strategies $a_{i,t}$, the capital of firm $i$ as well as competitors. Models of oligopoly with investment and firm entry and exit have been studied in \citet{ericson1995markov}. And in that framework, multiple equilibira are commonly observed, as proved in \citet{doraszelski2010computable}.
The concept of {\em experience-based equilibrium} was introduced in \citet{FershtmanPakes}, with possibly asymmetric information. Hence, firms use past payoffs to reinforce the probability of choosing an action. In that framework, agents explicitly construct beliefs, which is no longer necessary with reinforcement learning.

With adaptative learning, \citet{MarcetSargent1,MarcetSargent2} proved that there was convergence to  a rational expectations equilibrium. The reinforcement learning model is here similar to the previous one, there are no assumption about belief of opponents' strategies. Somehow, those algorithms are more related to evolutionary games.
\citet{Brown51} suggested that firms could form beliefs about competitors' choice probabilities, using some {\em fictitious plays}, also called {\em Cournot learnning} (studied more deeply in \citet{Hopkins}). \citet{Bernheim84} and \citet{Pearce84} added assumptions on firms beliefs, called {\em rationalizability}, under which we can end-up with Nash equilibria.

\citet{maskin.tirole:theory1,maskin.tirole:theory2} considered the case where two firms compete in a Stackelberg competition: they alternate in moving, and then commit to a price for two periods, before (possibly) adjusting. They did observe cycles and tacit collusion within the two firms. Such a result was confirmed by \citet{Kimbrough2008} and \citet{WaltmanKaymak}. The later studied repeated Cournot games where all players act simultaneously. They study the use of $Q$-learning for modeling the learning behavior of firms in that repeated Cournot oligopoly games, and they show that $Q$-learning firms generally learn to collude with each other, although full collusion usually does not emerge. Such a behavior was also observed in \citet{Schwalbe} where self-learning price-setting algorithms can coordinate their pricing behavior to achieve a collusive outcome that maximizes the joint profits of the firms using them.

\subsection{Applications in Finance}\label{sec:finance}

The dynamic control or hedge of risks on financial markets is a natural playground for the use of reinforcement learning algorithms. In the literature, dynamic risk management problems have been extensively studied in model-driven settings, using the tools from dynamic programming either in continuous or discrete time. In such framework, reinforcement learning algorithms naturally opens the door to innovative model-free numerical approximation schemes for hedging strategies, as soon as a realistic financial market simulator is available. Such simulator may typically incorporate market imperfections and frictions  (transaction costs, market impact, liquidity issues...). In the following sections, we detail more specifically recent applications on three topics of interest in such context: pricing and hedging of financial derivatives, optimal asset allocation and market impact modeling. 

\subsubsection{Risk management}

The valuation and hedging of financial derivatives are usually tackled in the quantitative finance literature using model-driven decision rules in a stochastic environment. Namely, for given model dynamics of the assets on a financial market, pricing and hedging of a derivative boils down to solving a dynamic optimal control problem for a well chosen arbitrage free martingale measure. The practical hedging strategy then makes use of the so-called Greeks, the sensitivities of the risk valuation to the different parameters of the model.

Such analysis usually lacks efficient numerical approximation methods in high dimensional settings, as well as precise tractable analytical solutions in the presence of realistic market frictions or imperfections. In the spirit of \cite{weinan2017deep} , \cite{buehler2019deep} introduced the idea of using reinforcement learning based algorithm in such context, see also \cite{fcamp2019risk}. Let consider given a realistic simulator of the financial market possible trajectories. We can encompass the price and/or hedging strategy of the financial derivative in a neural deep network (or any other approximating class of function), and train/estimate the approximating function in a dynamic way. At each iteration, we measure the empirical performance (i.e. loss) of the hedging strategy obtained on a large number of Monte Carlo simulations, and update its parameters dynamically using any typical reinforcement learning algorithm.  In particular, such approach allows to encompass scalable high dimensional risk dynamics as well as realistic market frictions or hedging using a large number of financial derivatives. 

The design of the market simulator of course requires model-driven assumptions, such as the choice of a particular class of volatility models, as well as its calibration. Nevertheless, we can mention recent attempts on the design of model free financial market simulator based on generative methods, such as the one developed e.g. in \cite{wiese2019deep,wiese2019quant}.

\subsubsection{Portfolio allocation}
In a similar manner, the design of dynamic optimal investment strategy naturally falls into the scope of  reinforcement learning type algorithms. Such observation goes back to \cite{935097} and has developed a growing interest in the recent literature  \cite{deng2016deep, Almahdi2017AnAP}:  Classical Mean-variance trade-off in a continuous time setting is for example revisited in \cite{wang2019continuous} using such viewpoint. Being given a financial market simulator together with choices of return and risk measurement methods written in terms of running or terminal rewards, one can learn optimal investment strategies using typical reinforcement learning algorithms.  

One could argue that such algorithms for portfolio allocation may often be reduced to less sophisticate online or bandit type learning algorithms \cite{li2014online}. 
Such argumentation does not remain valid in the more realistic cases where the investor has a significant impact on the financial assets dynamics, as discussed in the next section.

\subsubsection{Market microstructure}

When trades occur at a very high frequency or concern a large volume of shares, buying and selling orders have an impact on the financial market evolution, that one can not neglect. It modifies the shape of the order book, containing the list of waiting orders chosen by the other traders of the market. Being given a realistic order book dynamics simulator (or using the financial market as such), one can optimize using Reinforcement Learning algorithms the  dynamic use of market and limit orders, see \cite{spooner2018market,gueant2020deep, baldacci2019market}. The environment is given by the current order book shapes while the state typically represents the inventory of the trader, on a possibly high-dimensional financial market. 

Such framework is with no doubt a perfect fit for reinforcement learning algorithms. Nevertheless, a finer modeling perspective should take into account that the order book dynamics result from the aggregation of other traders actions, i.e. buy or sell orders. Hence, as observed e.g. in \cite{ganesh2019reinforcement, vyetrenko2019risk}, such setting is more precisely described as a multi-agent learning problem, as the one described above in Section \ref{Sec_Games}.

The practical use of reinforcement based learning algorithms on financial markets suffers two main drawbacks. The first one is the difficulty to create a realistic financial market simulator, together with the necessity to create a robust optimal trading strategy, in response to the differences between the real market and the virtual one. The second and main one is the lack of stationarity of the financial dynamics, which hereby do not allow to apply efficiently on future market dynamics, the investment strategies learned on the past market data points. Besides, the aggregate use of model-free  approaches combined with hardly interpretable black box output policy shall inevitably lead to hardly controllable financial market dynamics.

\section{Conclusion}

Deep Reinforcement learning is nowadays the most popular technique for (artificial) agent to learn closely optimal strategy by experience. Majors companies are training self driving cars using reinforcement learning (see \cite{Folkers_2019}, or \cite{kiran2020deep} for a state-of-the-art). Such techniques are extremely powerful to models behaviors of animals, consumers, investors, etc. Economists have laid the groundwork for this literature, but computational difficulties slowed them down. Recent advances in computational science are extremely promising, and complex economic or financial problems would benefit from being reviewed in the light of these new results.

Nevertheless, algorithms perform well assuming that a lot of information is available. More importantly, as the exploration may represent a very large number of possibilities, the use of deep reinforcement learning algorithms rapidly requires very important computer power. In finance, despite the lack of stationary of the market, it is worth noting that these algorithms begin to be quite popular.

\bibliography{bibliography}

\end{document}